\documentstyle[12pt]{article}
\def\A{{\cal A}}                           
\def\B{{\cal B}}
                           
\def\D{{\cal D}}
\def\1{\parallel 1>}
\def\T{\tilde{T}}
\def\s{\sinh}
\def\c{\cosh}

\begin{document}

\begin{titlepage}
\title{Exact solution of the $SU_{q}(n)$ invariant quantum spin chains}

\author{H.J. de Vega\\
A. Gonz\'alez--Ruiz\thanks{Permanent address: Departamento
de F\'{\i}sica Te\'orica, Universidad Complutense, 28040 Madrid,
ESPA\~NA} \\
{\it L.P.T.H.E.} \\
{\it Tour 16, 1er \'etage, Universit\'e Paris VI} \\
{\it 4 Place Jussieu, 75252 Paris cedex 05, FRANCE}}
\date{}
\maketitle

\begin{abstract}
The Nested Bethe Ansatz is generalized to open boundary conditions.
 This is used to find the exact eigenvectors
and eigenvalues of the $A_{n-1}$ vertex model with fixed open boundary
conditions and the corresponding $SU_{q}(n)$ invariant hamiltonian.
 The Bethe Ansatz
equations obtained are solved in the thermodynamic limit giving the
vertex model free energy and the hamiltonian ground
state  energy including the corresponding boundary contributions. 
\end{abstract}

\vskip-16.0cm
\rightline{}
\rightline{{\bf LPTHE--PAR 93/38}}
\rightline{{\bf June 1993}}
\vskip2cm  

\end{titlepage}

\begin{section}{Introduction}

\indent The nested Bethe Ansatz (NBA) is probably the most sophisticated
algebraic construction of eigenvectors for integrable lattice models. 
It appears
in models where the underlying quantum group is of rank larger than
one. 

In the context of the algebraic Bethe Ansatz \cite{ft} the NBA for the 
 $A_{n-1}$ trigonometric and hyperbolic vertex model is given in
\cite{hr} where eigenvectors and eigenvalues are obtained for periodic
boundary conditions (PBC). In ref.\cite{kr} the NBA for the $Sp(2N)$ 
symmetric vertex model is given and in ref.\cite{mk} the NBA for $O(2N)$
 symmetric vertex model is constructed (always with PBC). Although the
NBA equations has been proposed for all Lie algebras \cite{krb} for
PBC, no general construction is yet available  for the
corresponding eigenvectors.

For fixed boundary conditions, the  algebraic Bethe Ansatz is known
for the six vertex model \cite{sk} and for the susy t-J model  \cite{fk}.

We present in this article the NBA construction of eigenvectors
and eigenvalues for the  $A_{n-1}$ trigonometric and hyperbolic vertex
model transfer matrix in the fundamental representation with fixed 
($SU_q(n)$ invariant) boundary conditions (b. c.).
That is, boundary conditions  determined by matrices
$K^{\pm}$ which satisfy the integrability condition together with
$R(\theta)$ \cite{sk},\cite{ks},\cite{mn}. 

The NBA is necessary to solve vertex models associated to Lie 
Algebras with rank $n-1 > 1$. 
For the six-vertex ($A_1$) model, the algebraic Bethe Ansatz gives the
transfer matrix eigenvectors as products of creation operators  
of pseudoparticles $\B(\theta)$ acting on the ferromagnetic ground state. 
When the  rank $n-1$ of the  associated to Lie  Algebra is  $n-1 > 1$,
one finds more than one creation
operator for pseudoparticles : $\B_a(\theta) ,  [2 \leq a \leq n]$.
Hence, as Bethe Ansatz for the transfer matrix eigenvectors,
a linear combination of $\B_a$'s acting on a  ferromagnetic ground state
 state and summed over the indices $a$ is proposed. Then one should
find the coefficients in such linear combination from the eigenvalue
condition. Surprisingly enough, these coefficients turn to obey
an eigenvector problem analogous to the original one
but with a new transfer matrix. This new transfer matrix is 
built from statistical weights obtained from the original ones
deleting the first row and column. This procedure can be iterated
as many times as necessary till one arrives to a one-by-one
transfer matrix. Then the problem  is solved in the sense that
reduces to a set of algebraic equations : the nested Bethe Ansatz 
equations (NBAE).

The use of fixed  boundary conditions seriously complicates
the resolution task. First, the commutation relations between
the pseudoparticle operators  $\B_a(\theta)$ and the transfer matrix
are much more involved than for PBC and generate therefore many more terms
when the transfer matrix is applied on the NBA vectors. Second,
the structure of the unwanted terms generated then is much richer.
There appear new algebraic identities that were trival in the periodic
  case and  have to be proved now [see eq. (\ref{ciclo})].

In sec. 2 we review the   $A_{n-1}$ trigonometric and hyperbolic vertex
model with fixed b. c. and its associated $SU_q(n)$ invariant spin
chain. In sec. 3 we present the NBA construction of eigenvectors
for this model when fixed boundary consitions are chosen 
and derive the NBAE. This purely algebraic 
construction is valid for lattices
of arbitrary size $N$. We try to keep our presentation as pedagogical
as possible. Some calculations are given in the Appendices.
In sec. 4 we solve the NBAE in
the thermodynamic limit. We explicitly find the contribution to the 
free energy of the  $A_{n-1}$ trigonometric and hyperbolic vertex
model due to  the presence of the boundaries. From it, we derive the boundary
contribution to the ground state  energy of the $SU_q(n)$ invariant
hamiltonian. 

\end{section}

\begin{section}{Construction of the $SU_{q}(n)$ invariant spin chain}
The nonzero elements of the $A_{n-1}$ $R(\theta)$-matrix in the
fundamental representation can be written for the ferromagnetic
regime as:

\begin{eqnarray}
R^{ab}_{ab}(\theta)&=&\frac{\sinh\gamma}{\sinh(\theta+\gamma)}
e^{\theta\,{\rm sign}(a-b)}\;\;,\;\;a\neq b\nonumber\; \; ; \\
R^{ab}_{ba}(\theta)&=&\frac{\sinh\theta}{\sinh(\theta+\gamma)}\;\;, \;
\; a\neq b \; \; ;
\label{mran}\\
R^{aa}_{aa}(\theta)&=&1\nonumber\\
&&1\leq a,b\leq n\nonumber
\end{eqnarray}
 
All other elements are zero. For $n=2$, eq.(\ref{mran}) reduces
to the six vertex R-matrix up to a gauge transformation \cite{hr}.

The weights in the antiferromagnetic and gapfull regime follow from
eq.(\ref{mran}) upon replacing $\gamma\rightarrow -\gamma+i\pi$.

In the gapless and antiferromagnetic regime the R-matrix takes the
form:

\begin{eqnarray}
R^{ab}_{ab}(\theta)&=&\frac{\sin\gamma}{\sin(\gamma-\theta)}
e^{i\theta \,{\rm sign}(a-b)}\;\; ,\;\; a\neq b\nonumber \quad ;\\
R^{ab}_{ba}(\theta)&=&\frac{\sin\theta}{\sin(\gamma-\theta)}\;\;,\;\;a\neq
b \quad ; \label{mrtri}\\
R^{aa}_{aa}(\theta)&=&1\nonumber\\
&&1\leq a,b\leq n\nonumber
\end{eqnarray}

Gapfull and gapless antiferromagnetic regimes are related by
the transformation: $\gamma\rightarrow i\gamma,~ \theta \rightarrow
i\theta$.

In all regimes, $R(\theta)$ fullfils the Yang-Baxter equation:

\begin{eqnarray}
[1\otimes R(\theta-\theta^\prime)][R(\theta)\otimes 1][1\otimes
R(\theta^\prime)]\nonumber\\
=[R(\theta^\prime)\otimes 1][1\otimes
R(\theta)][R(\theta-\theta^\prime)\otimes 1]
\label{yb}
\end{eqnarray}

 The $R$ -matrix (\ref{mran}) does not enjoy $P$ and $T$ symmetry but just $PT$
invariance. It is not crossing invariant either but it obeys the
weaker property \cite{n,mn}:

\begin{equation}
\left[\left\{\left[S_{12}(\theta)^{t_{2}}\right]^{-1}\right\}^
{t_{2}}\right]^{-1}
= 
L(\theta,\gamma) M_{2} S_{12}(\theta +2\eta) M_{2}^{-1}
\label{sm}
\end{equation}

where $S=PR$ ($P^{ij}_{kl}=\delta^{i}_{l}\delta^{j}_{k}$) and  
$\eta$, $L$, $M$ are given by \cite{ks}:

\begin{eqnarray}
\eta & = & \frac{n}{2} \gamma\nonumber\\
M_{ab} & = & \delta_{ab} \  e^{(n-2a+1)\gamma} \hspace{2cm} 1 \leq a,b \leq n
\nonumber \\
L(\theta,\gamma) & = & \frac{\sinh(\theta+\gamma) \sinh[\theta + (n-1)
\gamma]}{ \sinh(\theta) \sinh(\theta + n \gamma)} \label{M} \nonumber \\
\end{eqnarray}

Also this R-matrix obeys:

\begin{equation}
R(\theta)R(-\theta)=1
\label{norma}
\end{equation}
 
We will consider in this paper boundary conditions defined by the 
K-matrices \cite{ks}:
 
\begin{eqnarray}
K^{+}_{ab}(\theta)&=&e^{(n-2a+1)\gamma}\frac{\s(2\theta+\gamma)}
{\s(2\theta+n\gamma)}\delta_{ab}\label{kp}\\
K^{-}_{ab}(\theta)&=&\delta_{ab}\label{km}\\
&&1\leq a,b\leq n\nonumber
\end{eqnarray}

for the right and left boundaries, respectively. They are solutions of
the equations \cite{sk,mn}:

\begin{eqnarray}
R(\theta-\theta^\prime)[K^{-}(\theta)\otimes
1]R(\theta+\theta^\prime)[K^{-}(\theta^\prime)\otimes 1]\nonumber\\ 
=[K^{-}(\theta^\prime)\otimes 1]R(\theta+\theta^\prime)[K^{-}(\theta)\otimes
1]R(\theta-\theta^\prime)
\label{sk1}
\end{eqnarray} 

\begin{eqnarray}
R(\theta-\theta^{\prime})K^{+}_{1}(\theta^{\prime})^{t_{1}} 
M_{1}^{-1} R(-\theta -\theta^{\prime} -2\eta)
K^{+}_{1}(\theta)^{t_{1}}M_{2}\nonumber\\
=K^{+}_{1}(\theta)^{t_{1}} M_{2} R(- \theta -\theta^{\prime} -2\eta)
M_{1}^{-1} K_{1}^{+}(\theta^{\prime})^{t_{1}}  R(\theta
-\theta^{\prime})
\label{rm}
\end{eqnarray}

Notice that the solutions to these equations can be multiplied by an
arbitary function of $\theta$. These functions were chosen in equations
(\ref{kp}), (\ref{km}) in order to have the term proportional to $\A$
in the transfer matrix with coefficient 
equal to 1 (see eq.(\ref{trad})).\\
The Yang-Baxter operators $T_{ab}(\theta,\tilde{\omega})$ are defined as usual:

\begin{equation}
T_{ab}(\theta,\tilde{\omega})=
\sum_{a_{1},\ldots,a_{N-1}}t_{a_{1}b}(\theta+\omega_{N})\otimes
t_{a_{2}a_{1}}(\theta+\omega_{N-1})\otimes\ldots\otimes
t_{aa_{N-1}}(\theta+\omega_{1})
\label{deft}
\end{equation}

where N is the number of sites,
 $\tilde{\omega}=
(\omega_{N},\omega_{N-1},\ldots,\omega_{1})$ 
and $\omega_{i}$ $(1\leq i\leq N)$ are
arbitrary inhomogeneities. These operators obey the relation:

\begin{equation}
R(\theta-\theta')[T(\theta,\tilde{\omega})\otimes T(\theta',\tilde{\omega})]=
[T(\theta',\tilde{\omega})\otimes T(\theta,\tilde{\omega})]R(\theta-\theta')
\label{yb1}
\end{equation}

The row to row transfer matrix for periodic boundary conditions is
given by:

\begin{equation}
\tau(\theta,\tilde{\omega})=\sum_{a}T_{aa}(\theta,\tilde{\omega})
\label{deftp}
\end{equation}

For fixed boundary conditions described by the matrices $K^{\pm}(\theta)$, one
uses the Yang-Baxter operators $U_{ab}(\theta,\tilde{\omega})$ defined
by\cite{sk}:

\begin{equation}
U_{ab}(\theta,\tilde{\omega})=\sum_{cd}T_{ac}(\theta,\tilde{\omega})
K^{-}_{cd}(\theta)T^{-1}_{db}(-\theta,\tilde{\omega})
\label{defu}
\end{equation}

Here $T^{-1}_{cb}(\theta,\tilde{\omega})$ is the inverse in both the
horizontal and vertical spaces. That is:

\begin{equation}
\sum_{b}T_{ab}(\theta,\tilde{\omega})T^{-1}_{bc}(\theta,\tilde{\omega})=1
\;\delta_{ac}
\end{equation}

Where $1$ is the identity in the vertical space.\\
The YB operators $U_{ab}(\theta,\tilde{\omega})$ fulfil the YB
algebra:

\begin{eqnarray}
R(\theta-\theta^\prime)[U(\theta,\tilde{\omega})\otimes
1]R(\theta+\theta^\prime)[U(\theta^\prime,\tilde{\omega})\otimes 1]\nonumber\\
=[U(\theta^\prime,\tilde{\omega})\otimes 1]R(\theta+\theta^\prime)
[U(\theta,\tilde{\omega})\otimes 1]R(\theta-\theta^\prime)
\label{ybau}
\end{eqnarray}

The fixed boundary condition transfer matrix is defined as:

\begin{equation}
t(\theta,\tilde{\omega})=\sum_{ab}K^{+}_{ab}(\theta)
U_{ab}(\theta,\tilde{\omega})
\label{mtdf}
\end{equation}

\noindent where $\tilde{\omega}=
(\omega_{N},\omega_{N-1},\ldots,\omega_{1})$, (see figure A).
Thanks to eqs. (\ref{sm}), (\ref{rm}), (\ref{ybau}) and (\ref{mtdf})
 the $t(\theta,\tilde{\omega})$ form a one parameter commuting family:
                                      
\begin{equation}
[t(\theta,\tilde{\omega}),t(\theta^\prime,\tilde{\omega})]=0
\label{cmtt}
\end{equation}

Furthermore, these transfer matrices built with $K^{\pm}$ given by 
(\ref{kp})-(\ref{km}) commute
with the $SU_{q}(n)$ generators as shown in refs. \cite{ks,mn}.\\
\indent The $SU_{q}(n)$ invariant hamiltonian associated to this transfer
matrix is given by \cite{ks}:

\begin{eqnarray}
&&H =\nonumber \\  
&&\sum_{j=1}^{N-1} \{ \sum_{\begin{array}{c} \scriptsize r,s=1
\\
 \scriptsize r > s \end{array}}^{n}  ( \prod_{l=s}^{r-1} \
(J_{l}^{+})^{(j)} \prod_{l=r-1}^{s} (J_{l}^{-})^{(j+1)} + 
\prod_{l=r-1}^{s} (J_{l}^{-})^{(j)} \prod_{l=s}^{r-1} \
(J_{l}^{+})^{(j+1)}) + \nonumber \\
&&\frac{\cosh\gamma}{n} [ \sum_{ \begin{array}{c} \scriptsize r,s=1
\\ \scriptsize r > s \end{array}}^{n-1} s(n-r)
(h_{r}^{(j)}h_{s}^{(j+1)} + h_{s}^{(j)}h_{r}^{(j+1)}) +
\sum_{r=1}^{n-1} r(n-r) h_{r}^{(j)}h_{r}^{(j+1)} ] + \nonumber
\\
&&\frac{\sinh\gamma}{n}\sum_{\begin{array}{c} \scriptsize r,s=1
\\
 \scriptsize r > s \end{array}}^{n-1} s(r-s)(n-r)(
h_{r}^{(j)}h_{s}^{(j+1)} - h_{s}^{(j)}h_{r}^{(j+1)}) \} +\nonumber \\  
&&\frac{\sinh\gamma}{n}\sum_{r=1}^{n-1} r(n-r) (
h_{r}^{(N)}-h_{r}^{(1)} )
\label{hsun} 
\end{eqnarray} 

where we have ommited a term proportional to the unity operator.
Here $N$ is the number of sites, $J_{l}^{+} \equiv  e_{l
l+1}$, $J_{l}^{-} \equiv e_{l+1 l}$ and $h_{l} \equiv e_{l l}-e_{l+1 l+1}$ 
are the
$SU(n)$ generators in the fundamental representation with $(e_{l
m})_{i j} \equiv  \delta_{l i}\delta_{m j}$. It is easily seen that this
hamiltonian coincides for $n=2$ with the $SU_{q}(2)$ invariant one,
discussed in \cite{ps}-\cite{qba1}.

\section{Nested Bethe Ansatz for the open $SU_{q}(n)$ invariant
transfer matrix}
In this section we give the NBA construction for the $A_{n-1}$ vertex
model with open boundary conditions.\\
To make contact with the known case $n=2$ 
is convenient to work with slightly modified local vertices:

\begin{equation}
[t_{ab}(\theta)]_{cd}=R^{bd}_{ca}(\theta-\gamma/2)
\label{vert}
\end{equation}

It is also convenient to introduce the notation:

\begin{eqnarray}
\A(\theta)&=&U_{11}(\theta)\nonumber\\
\B_{a}(\theta)&=&U_{1a}(\theta)
\label{abd}\\
\D_{ab}(\theta)&=&U_{ab}(\theta)\nonumber\\
&&2\leq a,b\leq n \nonumber
\end{eqnarray}

The Yang-Baxter algebra fulfilled by these operators follows by
inserting eqs. (\ref{mran}) and  (\ref{abd}) in eq.
 (\ref{ybau}) (see appendix A).\\
Actually it is more convenient to work with the operators:

\begin{eqnarray}
\hat{\D}_{bd}(\theta)&=&\frac{1}{\s(2\theta-\gamma)}[e^{2\theta-\gamma}
\s(2\theta) \;
\D_{bd}(\theta)-\s\gamma\;\delta_{bd}\; \A(\theta)]\label{c1}\\
\hat{\B}_{c}(\theta)&=&\frac{\s(2\theta)}{\s(2\theta-\gamma)}\B_{c}(\theta)
\label{c2}
\end{eqnarray}

The commutation relations are then given by:

\begin{eqnarray}
\A(\theta)\, \hat{\B}_{c}(\theta^\prime)&
=&\frac{\s(\theta+\theta^\prime-\gamma)
\s(\theta-\theta^\prime-\gamma)}{\s(\theta+\theta^\prime)
\s(\theta-\theta^\prime)}\;\hat{\B}_{c}(\theta^\prime)\,\A(\theta)\nonumber\\
&&+\frac{\s\gamma
e^{\theta-\theta^\prime}\s(2\theta-\gamma)}
{\s(2\theta) \s(\theta-\theta^\prime)}\;\hat{\B}_{c}(\theta) \,
\A(\theta^\prime)
\label{abba}\\
&&-\frac{\s\gamma
e^{\theta-\theta^\prime}\s(2\theta-\gamma)}{\s(2\theta)
\s(\theta+\theta^\prime)}\;\hat{\B}_{g}(\theta)\hat{\D}_{gc}
(\theta^\prime)\nonumber
\end{eqnarray}

\begin{eqnarray}
\hat{\D}_{bd}(\theta)\;\hat{\B}_{c}(\theta^\prime)
&=&\frac{\s(\theta+\theta^\prime+\gamma)\s(\theta-\theta^\prime+\gamma)}
{\s(\theta+\theta^\prime)\s(\theta-\theta^\prime)}\nonumber\\
&&\;R^{(2)}(\theta+\theta^\prime)^{eb}_{gh}\;R^{(2)}
(\theta-\theta^\prime)^{ih}_{cd}\;\hat{\B}_{e}(\theta^\prime)\,
\hat{\D}_{gi}(\theta)
\label{dbbd}\\
&&+\frac{\s\gamma
e^{\theta-\theta^\prime}\s(2\theta+\gamma)}{\s(\theta+\theta^\prime)
\s(2\theta)}
R^{(2)}(2\theta)^{gb}_{cd}\;\hat{\B}_{g}(\theta)\,\A(\theta^\prime)\nonumber\\
&&-\frac{\s\gamma
e^{\theta-\theta^\prime}\s(2\theta+\gamma)}{\s(\theta-\theta^\prime)
\s(2\theta)}
\;R^{(2)}(2\theta)^{gb}_{id}\;\hat{\B}_{g}(\theta)\,
\hat{\D}_{ic}(\theta^\prime)\nonumber
\end{eqnarray}

Where $R^{(2)}(\theta)^{ij}_{kl}$ is the original R matrix but with
indices $2\leq i,j,k,l\leq n$. 
The first term of the last equation may be seen as
the building block of a transfer matrix of a problem with $n-1$ states
per link, inhomogeneities $\theta'$ and local weights given by
$[t^{(2)}_{ab}(\theta)]_{cd}=R^{(2)}(\theta)^{bd}_{ca}$ with indices going
from 2 to n (notice  the change $\theta\rightarrow\theta+\gamma/2$ in
the local wheights from eq.(\ref{vert})). 
These commutation relations reduce to those of
\cite{qba1} for the case $n=2$ after cancelling the exponentials  by
an appropiate gauge transformation.\\
Our aim is to build eigenvectors of the transfer matrix
$t(\theta,\tilde{\omega})$ (defined by eq.(\ref{mtdf})). We find using eqs.
(\ref{mtdf}) and (\ref{c1}):

\begin{eqnarray}
t(\theta,\tilde{\omega})&=&\sum_{ab=1}^{n}K^{+}_{ab}(\theta-\gamma/2)
U_{ab}(\theta,\tilde{\omega})\nonumber\\
&=&\A(\theta)
+\frac{\s(2\theta-\gamma)}{\s(2\theta+(n-1)\gamma)}e^{-2\theta}\;
\sum^{n}_{a=2}e^{n-2(a-1)\gamma}\;\hat{\D}_{aa}(\theta)\nonumber\\
&=&\A(\theta)+\frac{\s(2\theta-\gamma)}{\s(2\theta+\gamma)}e^{-2\theta}\;
\sum^{n}_{a=2}K^{+(2)}_{aa}(\theta)\;\hat{\D}_{aa}(\theta)
\label{trad}
\end{eqnarray}

Where $K^{(2)+}(\theta)$ is   obtained from eq.(\ref{kp}) by making
$n\rightarrow n-1$ and reordering indices such that they run from 2 to n.
   $K^{(2)+}(\theta)$ is the $K^{+}$-matrix for the reduced problem
with local vertices  
$[t^{(2)}_{ab}(\theta)]_{cd}=R^{(2)}(\theta)^{bd}_{ca}\label{vert2}$.\\
It is easy to find an eigenstate of $t(\theta,\tilde{\omega})$,
 the so called reference state $\1$ given by:

\begin{equation}
\1=\otimes^{N}_{i=1}\left(
\begin{array}{c}
1\\0\\.\\.\\.\\0
\end{array}
\right)
\end{equation}

This ferromagnetic state is an eigenvector of both $\A(\theta)$ and
$\hat{\D}_{dd}(\theta)$ $(2\leq d\leq n)$ with eigenvalues:

\begin{eqnarray}
\A(\theta)\1&=&\1\nonumber\\
\hat{\D}_{dd}(\theta)\1&=&\Delta_{-}(\theta)\1
\label{autr}
\end{eqnarray}

 Where, see [appendix B]:

\begin{equation}
\Delta_{-}(\theta)=e^{2\theta}\prod_{i=1}^{N}\frac{\s(\theta+
\omega_{i}-\gamma/2)\s(\theta-\omega_{i}-\gamma/2)}{\s(\theta+\omega_{i}+
\gamma/2)\s(\theta-\omega_{i}+\gamma/2)}
\label{det}
\end{equation}

In addition, we find [see appendix B] that:

\begin{eqnarray}
\hat{\D}_{ij}(\theta)\1=0,\;i\neq j\nonumber\\
U_{a1}(\theta)\1=0,\;a\geq 2
\label{dsom}
\end{eqnarray}

Hence, only the $\hat{\B}_{a}(\theta)$'s acting on $\1$ give some
nonzero vector, not proportional to the $\1$ itself.\\
Therefore, in order to build generic eigenvectors we  repeatedly apply 
operators $\hat{\B}_{j}(\mu_{j})$ on the reference state $\1$ and
consider linear combinations. That is :

\begin{equation}
\Psi \equiv\sum_{2\leq i_{j}\leq
n}X^{i_{1}\ldots i_{r}}\hat{\B}_{i_{1}}(\mu_{1})\ldots
\hat{\B}_{i_{r}}(\mu_{r})\1\nonumber\\
=\hat{\B}(\mu_{1})\otimes\ldots\otimes\hat{\B}(\mu_{r})X\1
\label{bare}
\end{equation}

Here $\mu_{1}\ldots\mu_{r}$ and $X^{i_{1}\ldots i_{r}}$ are arbitrary
numbers. They will be constrained by requiring  $\Psi$ to be an
eigenvector of $t(\theta,\tilde{\omega})$. We can assume $\Psi$ to be
$\theta$ independent thanks to eq.(\ref{cmtt}).\\
\indent
 Our strategy goes as follows. Since $t(\theta,\tilde{\omega})$ is a
linear combination of $\A(\theta)$ and $\hat{\D}_{aa}(\theta)$ (see
eq.(\ref{trad})) we can apply separately each operator to $\Psi$.
Then we will use the commutation rules (\ref{abba}) and (\ref{dbbd}) to
push the operators $\A(\theta)$ and $\hat{\D}_{aa}(\theta)$ through
the $\hat{B}_{i_{j}}(\mu_{j})$ till $\A$ and $\hat{\D}_{ab}$ reach
$\1$. We use then eqs.(\ref{autr}) and (\ref{dsom}). Many terms
arise in this way. they can be classified in two types: wanted and
unwanted.\\
Wanted terms are those containing the original vectors: 

\begin{equation}
\hat{\B}_{i_{1}}(\mu_{1})\ldots\hat{\B}_{i_{r}}(\mu_{r})\1
\end{equation}

Unwanted terms are those where some argument $\mu_{j}$ is replaced by
$\theta$. That is, terms arising from the second and third terms in
the eqs. (\ref{abba}) and (\ref{dbbd}). These terms are called
``unwanted'' since they can never be proportional to $\Psi$ (here $\theta$
is an arbitrary complex number).\\
The wanted term in $\A(\theta)\Psi$ easily follows by repeatedly using
the first term in eq.(\ref{abba}). We have:

\begin{equation}
wanted\; term\; in\;
\A(\theta)\Psi=\prod_{j=1}^{r}\frac{\s(\theta+\mu_{j}
-\gamma)\s(\theta-\mu_{j}-\gamma)}{\s(\theta+\mu_{j})
\s(\theta-\mu_{j})}\Psi
\label{w1}
\end{equation}
 
An unwanted term where $\hat{\B}(\theta)$ replaces $\hat{\B}(\mu_{1})$
follows by using the second term in the rhs of (\ref{abba}) when
commuting $\A(\theta)\hat{\B}_{i_{1}}(\mu_{1})$ and the first term in
(\ref{abba}) for the subsequent commutations
$\A(\theta)\hat{\B}_{i_{j}}(\mu_{j})$ $(2\leq j\leq r)$. We find:

\begin{eqnarray}
\frac{\s\gamma \s(2\theta-\gamma)
e^{\theta-\mu_{1}}}{\s(2\theta)\s(\theta-\mu_{1})}
\prod^{r}_{j=2}\frac{\s(\mu_{1}+\mu_{j}-\gamma)\s(\mu_{1}-\mu_{j}-\gamma)}
{\s(\mu_{1}+\mu_{j})\s(\mu_{1}-\mu_{j})}\nonumber\\
\hat{\B}_{i_{1}}(\theta)\hat{\B}_{i_{2}}(\mu_{2})\ldots
\hat{\B}_{i_{r}}(\mu_{r})X^{i_{1}\ldots i_{r}}\1
\label{utau}
\end{eqnarray}

This calculation was rather simple because
 $\hat{\B}_{i_{1}}(\mu_{1})$ was the first
operator from the left. Now, we can find the other unwanted terms by
pushing the respective $\hat{\B}$'s to the left using the following
 cyclic symetry implied by eq.(\ref{bbbb}) in Appendix A:

\begin{eqnarray}
\hat{\B}(\mu_{1})\otimes\hat{\B}(\mu_{2})\otimes\ldots\otimes\hat{\B}(\mu_{r})
=\nonumber\\
\hat{\B}(\mu_{2})\otimes\hat{\B}(\mu_{3})\otimes\ldots\otimes\hat{\B}(\mu_{r})
\otimes\hat{\B}(\mu_{1})\;\tau^{(2)}(\mu_{1},\tilde{\mu})
\label{perm}
\end{eqnarray}

where:

\begin{eqnarray}
\tau^{(2)}(\theta,\tilde{\mu})=\sum^{n}_{a=2}T^{(2)}_{aa}(\theta,\tilde{\mu})
\end{eqnarray}

\noindent and $T^{(2)}_{aa}(\theta,\tilde{\mu})$ 
is given by eq.(\ref{deft})
with r sites,  indices $a_{i}$ running from 2 to n and local
weights $[t_{ab}^{(2)}(\theta)]_{ij}=R^{ja}_{bi}(\theta)$. That is,
$T^{(2)}_{ab}(\theta,\tilde{\mu})$ is the Yang-Baxter operator for a
restricted model with periodic boundary conditions, one less state
per link than in the original model and inhomogeneities
 $\tilde{\mu}=(\mu_{r},\ldots,\mu_{1})$ (see figure B). 
Notice that the inhomogeneities in this restricted model are
given by the parameters $\mu_{j}$ of the BA vectors (\ref{bare}).\\
{F}rom now on, as in formula (\ref{perm}), indices corresponding to
lines carrying identical inhomogeneities will be contracted,
 (see for example figure D).\\
Equation (\ref{perm}) tells us that the cyclic permutations
$\mu_{i}\rightarrow\mu_{i+1}$ followed by the action of
$\tau^{(2)}(\mu,\tilde{\mu})$ leaves $\Psi$ invariant. This property
obviously generalizes as :

\begin{eqnarray}
\hat{\B}(\mu_{1})\otimes\ldots\hat{\B}(\mu_{k})\otimes\ldots
\otimes\hat{\B}(\mu_{r})
=\nonumber\\
\hat{\B}(\mu_{k})\otimes\hat{\B}(\mu_{k+1})\otimes\ldots\otimes
\hat{\B}(\mu_{k-1})\; \tau^{(2)}_{k-1}\ldots\tau^{(2)}_{1}
\label{perg}
\end{eqnarray}

\noindent where $\tau^{(2)}_{j}=\tau^{(2)}(\mu_{j},\tilde{\mu})$.\\
Using this we can predict the form of the general unwanted
term where $\hat{\B}(\theta)$ replaces $\hat{\B}(\mu_{k})$ by looking
at  eqs. (\ref{utau})  and (\ref{perg}). We find:
 
\begin{eqnarray}
\frac{\s\gamma\s(2\theta-\gamma)\,e^{\theta-\mu_{k}}}{\s(2\theta)}
\; \sum^{r}_{k=1}\frac{e^{\theta-\mu_{k}}}{\s(\theta-\mu_{k})}\nonumber\\
\prod^{r}_{\begin{array}{c}\scriptsize j=1\\\scriptsize j\neq
k\end{array}}\frac{\s(\mu_{k}+\mu_{j}-\gamma)\s(\mu_{k}-\mu_{j}-\gamma)}
{\s(\mu_{k}+\mu_{j})\s(\mu_{k}-\mu_{j})}\nonumber\\
\hat{\B}(\theta)\otimes\hat{\B}(\mu_{k+1})\otimes\ldots
\otimes\hat{\B}(\mu_{r})\otimes\hat{\B}(\mu_{1})\otimes\ldots
\otimes\hat{\B}(\mu_{k-1})\label{summ}\\
\tau^{(2)}_{k-1}\ldots\tau^{(2)}_{1}X\1\nonumber
\end{eqnarray}

The third term in the right hand side of eq.(\ref{abba})
produces another kind of unwanted terms. (Notice that such terms are
absent for periodic boundary conditions \cite{hr}). 
We find from the third term in eq.(\ref{abba}) using then the
  first term
in eq.(\ref{dbbd}) $ (r-1) $ times and
 the preceding argument:

\begin{eqnarray}
-\frac{\s\gamma\s(2\theta-\gamma)}{\s 2\theta}\sum^{r}_{k=1}
\frac{e^{\theta-\mu_{k}}}{\s(\theta+\mu_{k})}
\Delta_{-}(\mu_{k})\nonumber\\
\prod^{r}_{\begin{array}{c}\scriptsize j=1\\\scriptsize j\neq k\end{array}}
\frac{\s(\mu_{k}+\mu_{j}+\gamma)\s(\mu_{k}-\mu_{j}+\gamma)}
{\s(\mu_{k}+\mu_{j})\s(\mu_{k}-\mu_{j})}
\nonumber\\
\hat{\B}(\theta)\otimes\hat{\B}(\mu_{k+1})
\otimes\ldots\otimes\hat{\B}
(\mu_{r})\otimes\hat{\B}(\mu_{1})\otimes\ldots\otimes\hat{\B}(\mu_{k-1})
\label{tunw}\\
t^{(2)}(\mu_{k};\bar{\mu})\tau^{(2)}_{k-1}\ldots
\tau^{(2)}_{1}\1X\nonumber
\end{eqnarray}  

\noindent where $t^{(2)}(\mu_{k};\bar{\mu})$ is a
transfer matrix like in eq.(\ref{mtdf}) but for a reduced model
with $n-1$ states per link, indices running from  2 to n, 
local weights given by
 $[t^{(2)}_{ab}(\theta)]_{cd}=R^{(2)}(\theta)^{bd}_{ca}$ and inhomogeneities
$\bar{\mu}=(\mu_{k-1},\ldots,\mu_{1},\mu_{r},\ldots,\mu_{k+1},\mu_{k})$.
 We have also used (see figure C) that:

\begin{eqnarray}
\sum_{d=2}^{n}R^{(2)}(2\theta)^{ed}_{gd}\;K^{(2)+}_{dd}(\theta)
=\delta_{ge}
\label{tr}
\end{eqnarray}
 
\noindent  and the cyclic symmetry argument
above (eq.(\ref{perg})). Notice that this term 
 is absent in the periodic case \cite{hr}.\\
This completes the analysis of $\A(\theta)\Psi$.\\
Let us now compute the action of (see eq.(\ref{trad})):

\begin{equation}
\frac{\s(2\theta-\gamma)}{\s(2\theta+\gamma)}e^{-2\theta}\;
\sum^{n}_{a=2} K^{+(2)}(\theta)_{aa}\hat{\D}_{aa}(\theta) 
\label{tab}
\end{equation}

\noindent on $\Psi$.\\
 As before, wanted and unwanted terms appear. The wanted
term follows by using repeatedly the first term in the right hand side
of eq.(\ref{dbbd}) when $\hat{\D}_{aa}(\theta)$ is commuted
through the $\hat{\B}(\mu_{j})$. We find:

\begin{eqnarray}
wanted\;term\;in\;\frac{\s(2\theta-\gamma)}{\s(2\theta+\gamma)}
\; e^{-2\theta} \;
\sum^{n}_{a=2} K^{+(2)}_{aa}(\theta)\hat{\D}_{aa}(\theta) \; \Psi=
\nonumber\\
e^{-2\theta}\;\frac{\s(2\theta-\gamma)}{\s(2\theta+\gamma)} \;
\prod^{r}_{j=1}\frac{\s(\theta+\mu_{j}+\gamma)\s(\theta-\mu_{j}+\gamma)}
{\s(\theta+\mu_{j})\s(\theta-\mu_{j})}\Delta_{-}(\theta)
\label{dwan}\\
\hat{\B}_{j_{1}}(\mu_{1})\ldots\hat{\B}_{j_{r}}(\mu_{r})\1t^{(2)}
(\theta;\tilde{\mu})^{j_{1}\ldots j_{r}}_{i_{1}\ldots i_{r}}
X^{i_{1}\ldots i_{r}}\nonumber
\end{eqnarray}

\noindent with 
$\tilde{\mu}=(\mu_{r},\ldots,\mu_{k+1},\mu_{k},\mu_{k-1},\ldots,\mu_{1})$.\\
We have collected the R-matrices from the first term in the right hand
side of (\ref{dbbd}) into $t^{(2)}(\theta;\tilde{\mu})$ as in the case
of eq.(\ref{tunw}) (notice the change
 $\theta-\gamma/2\rightarrow\theta$ with respect to
the original problem (\ref{vert})).\\
We see that this wanted term will be proportional to $\Psi$ if
the coefficients  $X^{i_{1}\ldots i_{r}}$ form an eigenvector of the
reduced transfer matrix $t^{(2)}(\theta;\tilde{\mu})$. That is, if we
require:

\begin{equation}
t^{(2)}(\theta;\tilde{\mu})X=\Lambda^{(2)}(\theta;\tilde{\mu})X
\label{reduc}
\end{equation}

The unwanted term coming from the second summand in eq.(\ref{dbbd})
follows by the usual symetry argument after using eq.(\ref{tr}) and the
first term of  eq.(\ref{abba}) $r-1$ times. This gives:

\begin{eqnarray}
\frac{\s\gamma\s(2\theta-\gamma)}{\s(2\theta)}\sum_{k=1}^{r}
\frac{e^{-\theta-\mu_{k}}}{\s(\theta+\mu_{k})}\nonumber\\
\prod^{r}_{\begin{array}{c}\scriptsize j=1\\\scriptsize j\neq
k\end{array}}\frac{\s(\mu_{k}+\mu_{j}-\gamma)\s(\mu_{k}-\mu_{j}-\gamma)}
{\s(\mu_{k}+\mu_{j})\s(\mu_{k}-\mu_{j})}\nonumber\\
\hat{\B}(\theta)\otimes\hat{\B}(\mu_{k+1})\otimes\ldots
\otimes\hat{\B}(\mu_{r})\otimes\hat{\B}(\mu_{1})\otimes\otimes\ldots
\otimes\hat{\B}(\mu_{k-1})\label{un1d}\\
\tau^{(2)}_{k-1}\ldots\tau^{(2)}_{1}X\1\nonumber
\end{eqnarray}

Notice again that  this term is absent
in the periodic case as it happens with eq. (\ref{tunw}).\\
The last term coming  from the action of eq. (\ref{tab}) in $\Psi$ 
follows  using the  third and first terms of eq.(\ref{dbbd})
 and the identity (\ref{tr}). We get :

\begin{eqnarray}
-\frac{\s\gamma\s(2\theta-\gamma)}{\s 2\theta}\sum^{r}_{k=1}
\frac{e^{-\theta-\mu_{k}}}{\s(\theta-\mu_{k})}
\Delta_{-}(\mu_{k})\nonumber\\
\prod^{r}_{\begin{array}{c}\scriptsize j=1\\\scriptsize j\neq k\end{array}}
\frac{\s(\mu_{k}+\mu_{j}+\gamma)\s(\mu_{k}-\mu_{j}+\gamma)}
{\s(\mu_{k}+\mu_{j})\s(\mu_{k}-\mu_{j})}
\nonumber\\
\hat{\B}(\theta)\otimes\hat{\B}(\mu_{k+1})
\otimes\ldots\otimes\hat{\B}
(\mu_{r})\otimes\hat{\B}(\mu_{1})\otimes\ldots\otimes\hat{\B}(\mu_{k-1})
\label{un2d}\\
t^{(2)}(\mu_{k},\bar{\mu})\tau^{(2)}_{k-1}\ldots
\tau^{(2)}_{1}\1X\nonumber
\end{eqnarray}  

with
$\bar{\mu}=(\mu_{k-1},\ldots,\mu_{1},\mu_{r},\ldots,\mu_{k+1},\mu_{k})$.\\
The sum of wanted terms reads from eqs. 
 (\ref{w1}), (\ref{dwan}) and (\ref{tred}):

\begin{eqnarray}
&&wanted\;term\;in\;t(\theta,\tilde{\omega})\Psi=\nonumber\\
&&[\prod^{r}_{j=1}\frac{\s(\theta+\mu_{j}-\gamma)\s(\theta-\mu_{j}-\gamma)}
{\s(\theta+\mu_{j})\s(\theta-\mu_{j})}\nonumber\\
&&+\frac{\s(2\theta-\gamma)}{\s(2\theta+\gamma)}\prod_{i=1}^{N}
\frac{\s(\theta+\omega_{i}-\gamma/2)
\s(\theta-\omega_{i}-\gamma/2)}
{\s(\theta+\omega_{i}+\gamma/2)\s(\theta-\omega_{i}+\gamma/2)}
\label{want}\\
&&\prod^{r}_{j=1}\frac{\s(\theta+\mu_{j}+\gamma)\s(\theta-\mu_{j}+\gamma)}
{\s(\theta+\mu_{j})\s(\theta-\mu_{j})}
\Lambda^{(2)}(\theta;\tilde{\mu})]\Psi\nonumber
\end{eqnarray}

\noindent where we have also  used eq.(\ref{det}). 
The term in brackets gives us the eigenvalue of the initial problem in
terms of $\Lambda^{(2)}(\theta;\tilde{\mu})$, the eigenvalue of the
reduced problem defined by $t^{(2)}(\theta,\tilde{\mu})$ with $n-1$
states per link and local weights
$[t^{(2)}_{ab}(\theta)]_{cd}=R^{(2)bd}_{ca}(\theta)$.\\
Before summing the unwanted terms we use the identity:

\begin{equation}
t^{(2)}(\mu_{k},\bar{\mu})\tau^{(2)}_{k-1}\ldots\tau^{(2)}_{1}=
\tau^{(2)}_{k-1}\ldots\tau^{(2)}_{1}t^{(2)}(\mu_{k},\tilde{\mu})
\label{ciclo}
\end{equation}

\noindent where
$\bar{\mu}=(\mu_{k-1},\ldots,\mu_{1},\mu_{r},\ldots,\mu_{k+1},\mu_{k})$
and
$\tilde{\mu}=(\mu_{r},\ldots,\mu_{k+1},\mu_{k},\mu_{k-1},\ldots,\mu_{1})$.\\
This identity tells us how to move cyclically the inhomogeneities in
the open chain. Although this is a trivial rotation in the periodic
case, this is not the case for open boundary conditions.
 The proof for three sites is in figure E, it
uses the Yang-Baxter eq.(\ref{yb}), the property (\ref{tr}) 
 and eq.(\ref{norma}).
Notice that it is enough to prove that:

\begin{equation}
t^{(2)}(\mu_{2},\bar{\mu})\tau^{(2)}_{1}=
\tau^{(2)}_{1}t^{(2)}(\mu_{2},\tilde{\mu})
\label{giro}
\end{equation}

\noindent with
$\bar{\mu}=(\mu_{1},\mu_{r},\ldots,\mu_{3},\mu_{2})$
and
$\tilde{\mu}=(\mu_{r},\ldots,\mu_{3},\mu_{2},\mu_{1})$, (see figure D).
 The proof for an arbitrary number of sites is straightforward using
repeatedly what was used for three sites.\\
Using this property one obtains  from eqs. (\ref{summ}), (\ref{tunw}),
(\ref{un1d}) and  (\ref{un2d}):

\begin{eqnarray}
&&\s(2\theta-\gamma)\s\gamma\sum_{k=1}^{r}\frac{1}{\s(\theta+\mu_{k})
\s(\theta-\mu_{k})}\nonumber\\
&&[\prod^{r}_{\begin{array}{c}\scriptsize j=1\\\scriptsize j\neq
k\end{array}}\frac{\s(\mu_{k}+\mu_{j}-\gamma)\s(\mu_{k}-\mu_{j}-\gamma)}
{\s(\mu_{k}+\mu_{j})
\s(\mu_{k}-\mu_{j})}\nonumber\\
&&-\prod^{r}_{\begin{array}{c}\scriptsize j=1\\\scriptsize j\neq k\end{array}}
\frac{\s(\mu_{k}+\mu_{j}+\gamma)
\s(\mu_{k}-\mu_{j}+\gamma)}{\s(\mu_{k}+\mu_{j})
\s(\mu_{k}-\mu_{j})}\nonumber\\
&&\prod_{i=1}^{N}\frac{\s(\mu_{k}+\omega_{i}-\gamma/2)
\s(\mu_{k}-\omega_{i}-\gamma/2)}
{\s(\mu_{k}+\omega_{i}+\gamma/2)\s(\mu_{k}-\omega_{i}+\gamma/2)}
\Lambda^{(2)}(\mu_{k};\tilde{\mu})]
\label{uwant}\\
&&\hat{\B}(\theta)\otimes\ldots\otimes\hat{\B}(\mu_{k-1})
\tau^{(2)}_{k-1}\ldots\tau^{(2)}_{1}X\1\nonumber
\end{eqnarray}

\noindent where we have used (\ref{reduc}).\\
The unwanted terms have to be zero if  $\Psi$ is to be an eigenvector.\\
In summary, we find the two following conditions to be satisfied :

\begin{eqnarray}
t^{(2)}(\theta;\tilde{\mu})\;X&=&\Lambda^{(2)}(\theta;\tilde{\mu})\;X
\label{tred}\\
\Lambda^{(2)}(\mu_{k};\tilde{\mu})&=&\prod_{i=1}^{N}
\frac{\s(\mu_{k}+\omega_{i}+\gamma/2)
\s(\mu_{k}-\omega_{i}+\gamma/2)}
{\s(\mu_{k}+\omega_{i}-\gamma/2)\s(\mu_{k}-\omega_{i}-\gamma/2)}\nonumber\\
&&\prod^{r}_{\begin{array}{c}\scriptsize j=1\\\scriptsize j\neq
k\end{array}}\frac{\s(\mu_{k}+\mu_{j}-\gamma)\s(\mu_{k}-\mu_{j}-\gamma)}
{\s(\mu_{k}+\mu_{j}+\gamma)\s(\mu_{k}-\mu_{j}+\gamma)}\label{polo}
\end{eqnarray}

It is easy to see that eq. (\ref{polo}) ensures the analyticity
of the wanted term  (\ref{want}) as a function of $\theta$
for $\theta = \pm\mu_j , 1\leq j \leq r$.\\
We have reduced the original problem of N sites, n states
per link and local weights given by  (\ref{vert}) to a problem of r
sites n-1 states per link and local weights 
 $[t_{ab}(\theta)]_{cd}=R^{(2)}(\theta)^{bd}_{ca}$ with
 inhomogeneities $\mu_{1}\ldots\mu_{r}$.\\
By analogy, we propose the following ansatz for the coeficients 
$X^{(1)}\equiv X^{i_{1}\ldots i_{r}}$ :

\begin{equation}
X^{(1)}=X^{(2)}\;\hat{\B}^{(2)}(\mu_{1}^{(2)},
\mu^{(1)})\otimes\ldots\otimes\hat{\B}^{(2)}(\mu_{p_{2}}^{(2)},
\mu^{(1)})\parallel 1^{(2)}>
\end{equation}

\noindent where $\parallel 1^{(2)}>=\otimes_{k=1}^{p_{1}}\1^{(k)}$ 
and $\1^{(k)}$ is
a $n-1$ component vector with the first component equal to one and the
rest vanishing, and 
 $\mu_{i}^{(1)} \equiv \mu_{i} , 1 \leq i \leq r\equiv p_{1}.$\\
This argument can be repeated as many times as necesary till 
the dimension of the vertical spaces reduce  to one. 
We get in this way a sequence of Bethe Ansatz, each of them contained
in the previous one. That is, a nested structure emerges.

It is important to
remark that the spectral parameter and the roots of the Bethe ansatz
suffer in the course of the construction a change 
$\theta\rightarrow\theta+\gamma/2$ from a Bethe Ansatz at a given level to
the next, and that the roots at each  level are
the inhomogeneities for the next level. This can be seen looking to the first
term of the commutation relations (\ref{comdb}) in Appendix A. 
Then one obtains:

\begin{eqnarray}
t^{(k+1)}(\theta,\tilde{\mu}^{(k)})X^{(k)}=
\Lambda^{(k+1)}(\theta,\tilde{\mu}^{(k)})X^{(k)}
\end{eqnarray}

\begin{eqnarray}
&&\Lambda^{(k)}(\theta,\tilde{\mu}^{(k-1)})=\prod_{j=1}^{p_{k}}
\frac{\s[\theta+\mu_{j}^{(k)}+(k-2)\gamma]\s(\theta-\mu_{j}^{(k)}-\gamma)}
{\s[\theta+\mu_{j}^{(k)}+(k-1)\gamma]\s(\theta-\mu_{j}^{(k)})}\nonumber\\
&&+\frac{\s[2\theta+(k-2)\gamma]}{\s(2\theta+k\gamma)}\nonumber\\
&&\prod_{j=1}^{p_{k-1}}
\frac{\s[\theta+\mu_{j}^{(k-1)}+(k-2)\gamma]\s(\theta-\mu_{j}^{(k-1)})}
{\s[\theta+\mu_{j}^{(k-1)}+(k-1)\gamma]\s(\theta-\mu_{j}^{(k-1)}+\gamma)}
\nonumber\\
&&\prod_{j=1}^{p_{k}}
\frac{\s(\theta+\mu_{j}^{(k)}+k\gamma)\s(\theta-\mu_{j}^{(k)}+\gamma)}
{\s[\theta+\mu_{j}^{(k)}+(k-1)\gamma]\s(\theta-\mu_{j}^{(k)})}
\Lambda^{(k+1)}(\theta,\mu^{(k)})
\label{autk}\\
&&1\leq k\leq n-1\;\;,\mu_{j}^{(0)}=\omega_{j}+\gamma/2
\;\;,\Lambda^{(n)}(\theta,\mu^{(n-1)})=1\nonumber
\end{eqnarray}

with $\mu^{(k)}_{i}$ obeying:

\begin{eqnarray}
&&\Lambda^{(k+1)}(\mu_{i}^{(k)},\tilde{\mu}^{(k)})=\nonumber\\
&&\prod_{j=1}^{p_{k-1}}
\frac{\s[\mu_{i}^{(k)}+\mu_{j}^{(k-1)}+(k-1)\gamma]
\s(\mu_{i}^{(k)}-\mu_{j}^{(k-1)}+\gamma)}
{\s[\mu_{i}^{(k)}+\mu_{j}^{(k-1)}+(k-2)\gamma]
\s(\mu_{i}^{(k)}-\mu_{j}^{(k-1)})}\nonumber\\
&&\prod^{p_{k}}_{\begin{array}{c} \scriptsize j=1\\\scriptsize j\neq
i\end{array}}\frac{\s[\mu_{i}^{(k)}+\mu_{j}^{(k)}+(k-2)\gamma]
\s(\mu_{i}^{(k)}-\mu_{j}^{(k)}-\gamma)}
{\s(\mu_{i}^{(k)}+\mu_{j}^{(k)}+k\gamma)
\s(\mu_{i}^{(k)}-\mu_{j}^{(k)}+\gamma)}
\label{polk}
\end{eqnarray}

Using the recurrence formula (\ref{autk}) we find for the eigenvalue 
of $t(\theta,\tilde{\omega})$ :

\begin{eqnarray}
&&\Lambda^{(1)}(\theta,\mu^{(0)})=\prod_{j=1}^{p_{0}}
\frac{\s(\theta+\mu^{(0)}_{j}-\gamma)\s(\theta-\mu_{j}^{(0)})}
{\s(\theta+\mu_{j}^{(0)})\s(\theta-\mu_{j}^{(0)}+\gamma)}\nonumber\\
&&\sum_{k=1}^{n}
\frac{\s(2\theta-\gamma)\s(2\theta)}
{\s[2\theta+(k-2)\gamma]\s[2\theta+(k-1)\gamma]}\nonumber\\
&&\prod_{j=1}^{p_{k-1}}
\frac{\s[\theta+\mu^{(k-1)}_{j}+(k-1)\gamma]\s(\theta-\mu^{(k-1)}_{j}+\gamma)}
{\s[\theta+\mu_{j}^{(k-1)}+(k-2)\gamma]\s(\theta-\mu_{j}^{(k-1)})}\nonumber\\
&&\prod_{j=1}^{p_{k}}
\frac{\s[\theta+\mu_{j}^{(k)}+(k-2)\gamma]\s(\theta-\mu_{j}^{(k)}-\gamma)}
{\s[\theta+\mu_{j}^{(k)}+(k-1)\gamma]\s(\theta-\mu_{j}^{(k)})}\label{autov}
\end{eqnarray}

\noindent Where in the last term the product over $p_{n}$ is substituted by
$1$.\\
\indent Let us now derive the Bethe Ansatz equations  for the parameters
$\mu_{i}^{(k)}$ $(1\leq i\leq p_{k}\;,\;1\leq k\leq n-1)$.\\
Changing $k$ by $k+1$ in eq.(\ref{autk}) and setting
$\theta=\mu_{i}^{(k)}$ yields:

\begin{eqnarray}
&&\Lambda^{(k+1)}(\mu_{i}^{(k)},\tilde{\mu}^{(k)})=\nonumber\\
&&\prod_{j=1}^{p_{k+1}}
\frac{\s[\mu_{i}^{(k)}+\mu_{j}^{(k+1)}+(k-1)\gamma]
\s(\mu_{i}^{(k)}-\mu_{j}^{(k+1)}-\gamma)}
{\s(\mu_{i}^{(k)}+\mu_{j}^{(k+1)}+k\gamma)\s(\mu_{i}^{(k)}-\mu_{j}^{(k+1)})}
\label{asa}
\end{eqnarray}

since the second term of  eq.(\ref{autk}) vanishes for this value of
$\theta$. Equating now eqs.(\ref{polk}) and (\ref{asa}):

\begin{eqnarray}
\prod^{p_{k}}_{\begin{array}{c}\scriptsize j=1\\\scriptsize j\neq
i\end{array}}\frac{\sin(\nu^{(k)}_{i}+\nu^{(k)}_{j}+i\gamma)
\sin(\nu^{(k)}_{i}-\nu^{(k)}_{j}+i\gamma)}
{\sin(\nu^{(k)}_{i}+\nu^{(k)}_{j}-i\gamma)
\sin(\nu^{(k)}_{i}-\nu^{(k)}_{j}-i\gamma)}
=\nonumber\\
\prod_{j=1}^{p_{k-1}}
\frac{\sin[\nu^{(k)}_{i}+\nu^{(k-1)}_{j}+i\gamma/2]\sin(\nu^{(k)}_{i}-
\nu^{(k-1)}_{j}+i\gamma/2)}
{\sin[\nu^{(k)}_{i}+\nu^{(k-1)}_{j}-i\gamma/2]\sin(\nu^{(k)}_{i}-
\nu^{(k-1)}_{j}-i\gamma/2)}\nonumber\\
\prod_{j=1}^{p_{k+1}}
\frac{\sin[\nu^{(k)}_{i}+\nu^{(k+1)}_{j}+i\gamma/2]\sin(\nu^{(k)}_{i}-
\nu^{(k+1)}_{j}+i\gamma/2)}
{\sin[\nu^{(k)}_{i}+\nu^{(k+1)}_{j}-i\gamma/2]\sin(\nu^{(k)}_{i}-
\nu^{(k+1)}_{j}-i\gamma/2)}
\label{equat}\\
1\leq i\leq p_{k}\;\;,\;\;1\leq k\leq n-1\nonumber
\end{eqnarray}

\noindent where we have set:

\begin{equation}
\mu^{(k)}_{j}=i\nu^{(k)}_{j}-(k-1)\gamma/2,
\;\;\;1\leq k\leq n-1,\;1\leq j\leq p_{k}
\label{numu}
\end{equation}

The function $\Lambda^{(1)}(\theta,\mu^{(0)})$ must not be singular at
the points $\theta=\mu^{k}_{j}$ $(1\leq j\leq p_{k}\;,\;1\leq k\leq
n-1)$ since the finite dimensional matrix $t(\theta,\tilde{\omega})$
is an analytic function of $\theta$. One can see that if the previous
Nested Bethe Ansatz Equations  (\ref{equat}) 
 are satisfied by the $\mu_{j}^{(k)}$ the
 residue of  $\Lambda^{(1)}(\theta,\mu^{(0)})$ at
$\theta=\mu_{j}^{(k)}$ and $\theta=-\mu_{j}^{(k)}-(k-1)\gamma$
vanish.\\
The NBAE for the gapless regime follow from eqs. (\ref{equat}) by
replacing $\gamma\rightarrow -i\gamma $, $\nu^{(k)}_{j}\rightarrow
i\nu^{(k)}_{j}$.

\end{section}

\begin{section}{Analysis of the Bethe Ansatz equations}

\indent In this section we investigate the solution of the NBAE associated to
the quantum group covariant NBA with the inhomogeneities at the first
level fixed to zero for simplicity.
 We will relate these equations with those of the periodic
case by means of the change of variables, (see \cite{qba1}):

\begin{eqnarray}
\lambda^{(k)}_{s}&=&\nu^{(k)}_{s}\nonumber\\
\lambda^{k}_{2P_{k}-s+1}&=&-\nu^{(k)}_{s}
\label{cam}\\
1\leq s\leq p_{k}&;&1\leq k\leq n-1\nonumber
\end{eqnarray}

In this way the NBAE can be written  for the gapless case 
as:

\begin{eqnarray}
&&\prod_{j=1}^{2p_{k}}\frac{\s(\lambda^{(k)}_{l}-\lambda^{(k)}_{j}+i\gamma)}
{\s(\lambda^{(k)}_{l}-\lambda^{(k)}_{j}-i\gamma)}=\nonumber\\
&&\frac{\s(\lambda^{(k)}_{l}+i\gamma/2)\s(\lambda^{(k)}_{l}-i(\pi-\gamma)/2)}
{\s(\lambda^{(k)}_{l}-i\gamma/2)\s(\lambda^{(k)}_{l}+i(\pi-\gamma)/2)}
\label{bae}\\
&&\prod_{j=1}^{2p_{k-1}}
\frac{\s(\lambda^{(k)}_{l}-\lambda^{(k-1)}_{j}+i\gamma/2)}
{\s(\lambda^{(k)}_{l}-\lambda^{(k-1)}_{j}-i\gamma/2)}\prod_{j=1}^{2p_{k+1}}
\frac{\s(\lambda^{(k)}_{l}-\lambda^{(k+1)}_{j}+i\gamma/2)}
{\s(\lambda^{(k)}_{l}-\lambda^{(k+1)}_{j}-i\gamma/2)}\nonumber\\
&&1\leq k\leq n-1;\;1\leq l\leq 2p_{k}\nonumber
\end{eqnarray}

[In the gapless regime, the statistical weights are given
by eq.(\ref{mrtri}) ]. \\
These equations are like the NBAE for periodic boundary conditions on
 2N sites with an additional source factor, (see \cite{hr}). In
addition we have the following constraints on the roots
$\lambda^{k}_{j}$:\\
(i) the total number of roots is even $(2p_{k})$ at every stage and
are symmetrically distributed with respect to the origin according to
eq. (\ref{cam}).\\
(ii) there is no root at the origin $\lambda^{(k)}=0$ at every stage
 due to the fact
that $\hat{\B}_{j}^{(k)}(\theta=-(k-1)\gamma/2)=0$ (see eq.
(\ref{c2})).\\
As usual, we take logarithms in eq.(\ref{bae}), yielding:

\begin{eqnarray}
&&\sum_{j=1}^{2p_{k+1}}\Phi(\lambda^{(k)}_{l}-\lambda^{(k+1)}_{j},\gamma/2)-
\sum_{j=1}^{2p_{k}}\Phi(\lambda^{(k)}_{l}-\lambda^{(k)}_{j},\gamma)\nonumber\\
&&+\sum_{j=1}^{2p_{k-1}}\Phi(\lambda^{(k)}_{l}-\lambda^{(k-1)}_{j},\gamma/2)
+\Phi(\lambda^{(k)}_{l},\gamma/2)-\Phi(\lambda^{(k)}_{l},(\pi-\gamma)/2)
\label{lbae}\\
&&=2\pi I^{(k)}_{l}\;\;\;1\leq k\leq n-1;\;1\leq l\leq 2p_{k}\nonumber
\end{eqnarray}

Where $I^{(k)}_{l}$ are integers and

\begin{equation}
\Phi(z,\gamma)=i\; \log\left[\frac{\s(i\gamma+z)}{\s(i\gamma-z)}\right]
\end{equation}

 We will consider the thermodinamic limit of eq.(\ref{bae}). One can now 
introduce a density of roots at every NBA level:

\begin{equation}
\rho^{(l)}(\lambda^{(l)}_{j})=\lim_{N\rightarrow\infty}
\frac{1}{N(\lambda^{(l)}_{j+1}-\lambda^{(l)}_{j}))}
\end{equation}

Defining now the counting functions as:

\begin{eqnarray}
Z^{(k)}_{N}(\lambda)\equiv\frac{1}{2\pi N}[\sum_{j=1}^{2p_{k+1}}
\Phi(\lambda-\lambda^{(k+1)}_{j},\gamma/2)-
\sum_{j=1}^{2p_{k}}\Phi(\lambda-\lambda^{(k)}_{j},\gamma)\nonumber\\
+\sum_{j=1}^{2p_{k-1}}\Phi(\lambda-\lambda^{(k-1)}_{j},\gamma/2)
+\Phi(\lambda,\gamma/2)-\Phi(\lambda,(\pi-\gamma)/2)]
\label{defz}\\
1\leq k\leq n-1\nonumber
\end{eqnarray}

And using that:

\begin{equation}
I^{(k)}_{j+1}-I^{(k)}_{j}=1+\sum_{h=1}^{N_{h}^{(k)}}\delta_{jj_{h(k)}}
\end{equation}

Where $N_{h}^{(k)}$ is the number of holes at level k. One can see 
for $N \to \infty $ that:

\begin{eqnarray}
\sigma^{k}(\lambda)\equiv\frac{dZ^{(k)}_{N}(\lambda)}{d\lambda}\approx
\frac{Z^{(k)}_{N}(\lambda_{j+1})-
Z^{(k)}_{N}(\lambda_{j})}{\lambda_{j+1}-\lambda_{j}}\nonumber\\
=\frac{1+\sum_{h=1}^{N_{h}^{(k)}}\delta_{jj_{h(k)}}}{(\lambda_{j+1}-
\lambda_{j})N}\approx
\rho^{(k)}(\lambda)+\frac{\delta(\lambda)}{N}+\frac{\sum_{h=1}^{N_{h}^{(k)}}
\delta(\lambda-\theta^{(k)}_{h})}{N}
\label{defs}
\end{eqnarray}

Where the term $\frac{\delta(\lambda)}{N}$ is produced by the
 hole at $\lambda=0$ at every NBA level.
In the limit of large N we have:

\begin{eqnarray}
\lim_{N\rightarrow\infty}\frac{1}{N}\sum_{j=1}^{2p_{k}}f(\lambda^{(k)}_{j})
=\int_{-\infty}^{\infty}d\lambda f(\lambda)\rho^{(k)}(\lambda)
\label{int}
\end{eqnarray}

Taking the derivative of eq.(\ref{defz}) with respect to
$\lambda$ and using eq. (\ref{defs}), we obtain integral equations for
$\sigma^{(k)}(\lambda)$. Let us start with the
 antiferromagnetic ground  state. That is, no holes besides 
$\lambda=0$, and no complex solutions. We find:

\begin{eqnarray}
&&\sigma^{k}(\lambda)-\sum_{m=1}^{n-1}\int_{-\infty}^{\infty}
d\mu K_{km}(\lambda-\mu)\sigma^{m}(\mu)\nonumber\\
&&=\frac{1}{2\pi N}\;\Phi^\prime(\lambda,\gamma/2)-\frac{1}{2\pi
N}\;\Phi^\prime(\lambda,(\pi-\gamma)/2)
\label{inteq}\\
&&+\frac{\delta_{k1}}{\pi}\;\Phi^\prime(\lambda,\gamma/2)-
\frac{1}{N}\sum_{m=1}^{n-1}K_{km}(\lambda)\nonumber
\end{eqnarray}

Where we set all the inhomogeneities equal to zero at the
first level ( $\omega_i=0\rightarrow\lambda^{(0)}_{i}=0$). 
 The kernel $K_{km}(\lambda)$ reads:

\begin{eqnarray}
2\pi K_{km}(\lambda)=\Phi^\prime(\lambda,\gamma/2)(\delta_{k,\;m+1}+
\delta_{k,\;m-1})-\Phi^\prime(\lambda,\gamma)\delta_{km}
\label{nucl}
\end{eqnarray}

This linear integral equation can be solved by means of the resolvent 
$R_{mn}(\lambda)$ given by the solution to the equation:

\begin{eqnarray}
\sum_{k=1}^{n-1}\int_{-\infty}^{\infty}R_{lk}(\tau-\lambda)[\delta_{km}
\delta(\lambda-\mu)-K_{km}(\lambda-\mu)]d\lambda=\delta(\tau-\mu)\delta_{lm}
\label{resol}
\end{eqnarray}

It is convenient to  Fourier transform  these quantities:

\begin{eqnarray}
R_{mn}(\lambda)=\int_{-\infty}^{\infty}\frac{dk}{2\pi}e^{ik\lambda}
\; \hat{R}_{mn}(k)
\label{four1}\\
\sigma^{l}(\lambda)=\int_{-\infty}^{\infty}\frac{dk}{2\pi}e^{ik\lambda}
\; \hat{\sigma}^{l}(k)
\label{four2}
\end{eqnarray}

The solution to eq. (\ref{resol}) is then given by \cite{hr}:

\begin{eqnarray}
\hat{R}_{ll^\prime}(2x)=\frac{\s(\pi x)\s[\gamma x(n-l_{>})]\s(\gamma
x l_{<})}{\s[x(\pi-\gamma)]\s(\gamma x n)\s(\gamma x)}
\label{resok}
\end{eqnarray}

\noindent where $l_{>}=max(l,l')$ and $l_{<}=min(l,l')$.
We then obtain for the derivative of the counting functions:

\begin{eqnarray}
\hat{\sigma}^{l}(k)&=&\frac{2\s[\gamma k (n-l)/2]}{\s(\gamma k n/2)}
 + 
\nonumber\\
&&+\frac{1}{N}-\frac{1}{N}(  \frac{2\s(k\gamma/4)
\c[k(\pi-\gamma)/4]}{\s(k\pi/2)} )
\sum_{m=1}^{n-1}\hat{R}_{lm}(k)
\label{sols}
\end{eqnarray}

One can see that this result reduces to the one given in \cite{qba1}
for the case $n=2$.\\
To compute the physically meanigful quantities only $\rho^{(1)}(k)$ is
needed. Using eqs. (\ref{defs})  and (\ref{sols})  is easy to see that:

\begin{eqnarray}
\hat{\rho}^{(1)}(k)&=&\frac{2\s[k\gamma(n-1)/2]}{\s(k\gamma n/2)}\\
&-&\frac{\s[k\gamma(n-1)/4]\c(k\pi/4)}{N\c(k\gamma n/4)\s[k(\pi-\gamma)/4]} 
\nonumber
\label{solr}
\end{eqnarray}

We have now the tools to evaluate the free energy of the  model
in the gapless regime. This is given by:

\begin{eqnarray}
f(\theta,\gamma,n)&
\begin{array}{c}
  \\
=\\
\scriptsize
N\rightarrow\infty
\end{array}&
-\frac{1}{N}\log\Lambda(\theta)
\nonumber\\
&\begin{array}{c}
  \\
=\\
\scriptsize
N\rightarrow\infty
\end{array} &
-\frac{i}{N} \sum_{j=1}^{2p_{1}}\Phi(i\theta-\lambda_{j},\gamma/2)\label{fre1}
\label{frea}\\
&=&-\int_{-\infty}^{\infty}\frac{dk}{k}e^{-k\theta}
\frac{\s[k(\pi-\gamma)/2]}{\s(k\pi/2)}\hat{\rho}^{(1)}(k)\label{fre2}
\end{eqnarray}

Note that we have made $\theta\rightarrow\theta+\gamma/2$, that is we
have returned to the local weights where $t(0,\tilde{\omega})\propto
1$.\\
Using  the expresion for $\hat{\rho}^{(1)}(k)$ (eq. (\ref{solr})) in 
(\ref{fre2}) the final result for the free energy is:

\begin{eqnarray}
&&f(\theta,\gamma,n)=4\int_{0}^{\infty}\frac{dx}{x}
\s(2x\theta)\frac{\s[x(\pi-\gamma)]\s[x\gamma(n-1)]}{\s(x\pi)
\s(x\gamma n)}
\label{free}\\
&&-\frac{2}{N}\int_{0}^{\infty}\frac{dx}{x}\s(2x\theta)
\frac{\c[x(\pi-\gamma)/2]\s[(n-1)x\gamma/2]}{\s(x\pi/2)\c(xn\gamma/2)}
\nonumber
\end{eqnarray}

The first term here is the known bulk free energy, (see \cite{hr}).
 The second term is the correction
produced by the open boundary conditions (that give quantum group
invariance).\\
The ground state energy for the $SU_{q}(n)$ invariant hamiltonian is
obtainined by using:

\begin{equation}
H=-\frac{\sin\gamma}{2}\dot{t}(0,0)+(N-1)\frac{(n-1)}{n}\cos\gamma
\end{equation}

We obtain, (deriving
(\ref{free}) with respect to $\theta$):

\begin{eqnarray}
e_{\infty}(\gamma)&=&\frac{n-1}{n}\cos\gamma-4\sin\gamma\int_{0}^{\infty}dx
\frac{\s[x(\pi-\gamma)]\s[x\gamma(n-1)]}
{\s(x\pi)\s(x\gamma n)}
\label{efund}\\
&-&\frac{(n-1)}{Nn}\cos\gamma
+\frac{2\sin\gamma}{N}\int_{0}^{\infty}dx
\frac{\c[x(\pi-\gamma)/2]\s[(n-1)x\gamma/2]}{\s(x\pi/2)\c(x\gamma n/2)}
\nonumber
\end{eqnarray}

In the special case $n = 2$, this reduces to the result in \cite{bq}.

The surface energy contribution in eq.(\ref{efund}) :
\begin{eqnarray}
e^S(\gamma)&=&-\frac{n-1}{n}\cos\gamma
+2\sin\gamma\int_{0}^{\infty}dx
\frac{\c[x(\pi-\gamma)/2]\s[(n-1)x\gamma/2]}{\s(x\pi/2)\c(x\gamma n/2)},
\label{enesup}
\end{eqnarray}

takes a simpler form in the $\gamma = 0$ (isotropic) limit.
We find
\begin{eqnarray}
e^S(0)&=&-\frac{n-1}{n} 
+2 \int_{0}^{\infty}dx
\frac{\exp[-x/2] \s[(n-1)x/2]}{\c(x n/2)},
\label{eniso}
\end{eqnarray}

This integral can be expressed in terms of elementary functions
\cite{tabrus}:

\begin{eqnarray}
e^S(0)&=&-\frac{n-1}{n} 
+\frac{2}{n} \{ \frac{\pi}{2 \sin(\pi/n)}  - \ln 2 \nonumber\\
&-&
\sum_{k=0}^{E(\frac{n-1}{2})} \cos\left(\frac{2k+1}{n}\pi \right)
\log\left[2 - 2 \cos\left(\frac{2k+1}{n}\pi\right) \right] \} 
\label{enisre}
\end{eqnarray}

Here $E(x)$ stands for integer part of $x$.

Let us finally consider the gapfull antiferromagnetic regime.
In this case the NBAE are solved by expanding in Fourier series,
since the NBAE roots are in the interval $(-\pi/2,+\pi/2)$.
We write  the density of roots as follows:
\begin{eqnarray}
\sigma^{l}(\lambda)=\sum_{m=-\infty}^{\infty}\frac{e^{2im\lambda}}{2\pi}
\; \; \hat{\sigma}^{l}(m)
\label{four3}
\end{eqnarray}
where $\sigma^{l}(\lambda)$ obeys a system of integral equations
analogous to eq.(\ref{inteq}):

\begin{eqnarray}
&&\sigma^{k}(\lambda)-\sum_{m=1}^{n-1}\int_{-\pi/2}^{+\pi/2}
d\mu K_{km}(\lambda-\mu)\sigma^{m}(\mu)\nonumber\\
&&=\frac{1}{2\pi N}\Phi^\prime(\lambda,\gamma/2)+\frac{1}{2\pi
N}\Phi^\prime(\lambda,(i\pi+\gamma)/2)
\label{intrig}\\
&&+\frac{\delta_{k1}}{\pi}\Phi^\prime(\lambda,\gamma/2)-
\frac{1}{N}\sum_{m=1}^{n-1}K_{km}(\lambda)\nonumber
\end{eqnarray}
where now,

\begin{equation}
\Phi(z,\gamma)=i\; \log\left[\frac{\sin(i\gamma+z)}{\sin(i\gamma-z)}\right]
\label{fitr}
\end{equation}
and the kernel $K_{km}(\lambda)$ is given by eq.(\ref{nucl}).
We find as  solution of eq.(\ref{intrig}):

\begin{eqnarray}
\hat{\sigma}^{l}(m)&=&\frac{4\s[\gamma m (n-l)]}{\s(\gamma m n)} + 
\nonumber\\
&&+\frac{1}{N}\left( 2 +\left\{ [ 1 + (-1)^m ]\exp(-|m| \gamma) - 1
\right\}
\sum_{k=1}^{n-1}\hat{R}_{lk}(m) \right)
\label{sotr}
\end{eqnarray}

Where $\hat{R}_{lk}(m)$ is the resolvent of eq.(\ref{intrig})
in Fourier space.
Then, using eq.(\ref{defs}), we find

\begin{eqnarray} 
\hat{\rho}^{1}(m)&=&\frac{4\s[\gamma m (n-1)]}{\s(\gamma m n)}  
+\frac{h_m}{N}\;\;,
\label{sotr1}
\end{eqnarray}

where

\begin{equation}
h_m \equiv  \frac{(-1)^m \s[(n-1)\gamma m/2] + \exp[-(n-1)\gamma
|m|/2] \s(\gamma m) }{\c[\gamma m n/2]\s(\gamma m/2)}\nonumber\\
\label{hache}
\end{equation}

We find upon inserting $\hat{\rho}^{1}(m)$ and $\Phi(z,\gamma)$
given by eqs.(\ref{fitr})-(\ref{sotr1}) in eq.(\ref{frea}) 

\begin{eqnarray}
f(\theta,\gamma,n)= 4 \theta ( 1 - \frac{1}{n} ) +
4 \sum_{m=1}^{\infty}\frac{e^{-m\gamma} \s[\gamma m (n-1)]
\s[2m\theta]}{m \s(\gamma m n)} \nonumber \\
+\frac{1}{N} \left[ (n+1) \theta + 
\sum _{m =1}^{\infty}\frac{e^{-m\gamma}h_m\s[2m\theta]}{m}
\right]
\label{enetri}
\end{eqnarray}

The first two terms correspond to  the known bulk 
free energy, (see \cite{hr}).
 The second term is the correction
produced by the open boundary conditions (that give quantum group
invariance).

\end{section}

\section{Conclusions}
We have presented the generalization of the Nested Bethe Ansatz to the
quantum group invariant case. It will be  interesting to generalize it to
the cases where the $K^{\pm}$ matrices are the general diagonal
solutions given in \cite{ks}.\\
It also  remains to study the quantum group  properties of the NBA
states as the highest weight property. Moreover, a rich structure 
must arise for the reduced models when $\gamma/\pi$ is a rational
number \cite{redu}. \\
It  would be interesting to study this construction for
algebras different to $A_{n-1}$. That is, to generalize the work in
refs. \cite{kr}-\cite{krb} to open boundary conditions

\begin{section}{Appendix A : commutation relations}
 We begin putting explicitly all indices in
eq. (\ref{sk1}). This yields  (from now on we will supose sum
over repeated indices):

\begin{equation}
\begin{array}{l}
M^{ab}_{cd}\equiv R(\theta-\theta^\prime)^{ab}_{ef}\;U_{eg}(\theta)R(\theta
 +\theta^\prime)^{gf}_{hd}\;U_{hc}(\theta^\prime)=\\
N^{ab}_{cd}\equiv U_{ae}(\theta^\prime)\;R(\theta +\theta^\prime)^{eb}_{fg}
\;U_{fh}(\theta)\;R(\theta - \theta^\prime)^{hg}_{cd}
\end{array}
\end{equation}

As we want to obtain the commutation relations between
$\A(\theta)=U_{11}(\theta)$, $\D_{bd}(\theta)=U_{bd}(\theta)$  and 
$\B_{c}(\theta)=U_{1c}(\theta)$ where ($ b,c,d \geq 2$ ) , we study
the equalities $ M^{11}_{1c}=N^{11}_{1c}$
and $M^{1b}_{cd}=N^{1b}_{cd}$. This gives:

\begin{eqnarray}
\A(\theta^\prime)\;\B_{c}(\theta)&=&\frac{\s(\theta+\theta')
\s(\theta-\theta'+\gamma)}
{\s(\theta-\theta^\prime)\s(\theta+\theta^\prime+\gamma)}
\;\B_{c}(\theta)\A(\theta^\prime)\nonumber\\
&&-\frac{e^{-(\theta-\theta^\prime)}\s(\theta+\theta^\prime)\s\gamma}
{\s(\theta-\theta^\prime)\s(\theta+\theta^\prime+\gamma)}
\;\B_{c}(\theta^\prime)\A(\theta)\nonumber
\\
&&-\frac{e^{(\theta+\theta^\prime)}\s\gamma}{\s(\theta+\theta'+\gamma)}
\;\B_{g}(\theta^\prime)\D_{gc}(\theta)
\label{ca1}
\end{eqnarray}

\begin{eqnarray}
\D_{bd}(\theta)\;\B_{c}(\theta^\prime)&=&\frac{\s(\theta+\theta'+\gamma)
\s(\theta-\theta'+\gamma)}{\s(\theta+\theta^\prime)
 \s(\theta-\theta^\prime)}\nonumber\\
&&\{ R^{(2)}(\theta+\theta^\prime)^{eb}_{gh}\;
R^{(2)}(\theta-\theta^\prime)^{ih}_{cd}\;\B_{e}(\theta^\prime)\;
\D_{gi}(\theta)\nonumber\\
&&-\frac{e^{-(\theta-\theta^\prime)}\s\gamma}{\s(\theta-\theta'+\gamma)}
R^{(2)}(\theta+\theta^\prime)^{gb}_{id}\;
\B_{g}(\theta)\;\D_{ic}(\theta^\prime)\nonumber\\
&&+\frac{e^{-(\theta+\theta^\prime)}\s\gamma}{\s(\theta+\theta'+\gamma)}
[R^{(2)}(\theta-\theta^\prime)^{ib}_{cd}\;
\A(\theta^\prime)\;\B_{i}(\theta)\nonumber \\
&&-\frac{e^{-(\theta-\theta^\prime)}\s\gamma}{\s(\theta-\theta'+\gamma)}
\A(\theta)\;\B_{c}(\theta^\prime)\;
 \delta_{bd}]\}
\label{cd1}
\end{eqnarray}

Where $R^{(2)}(\theta)^{ij}_{kl}$ is the original R matrix but with
indices $2\leq i,j,k,l\leq n$.
We would like to have all the $\B$'s to the left of the $\A$'s  in
the right hand side of eq.(\ref{cd1}). For that, one substitutes
eq. (\ref{ca1}) in the last two terms of eq. (\ref{cd1}). One
obtains in this way a long expresion that we omit.\\
To simplify the calculi of Bethe Ansatz we look now for a linear 
change of the operators such that no term proportional to
 $\B_{g}(\theta^\prime)\A(\theta)$ remains in the commutation of $\D$'s and
$\B$'s.
The most general linear change would be of the form:
\begin{equation}
\hat{\D}_{bd}(\theta)=\alpha^{rs}_{bd}(\theta)\D_{rs}(\theta)+
 \beta_{bd}(\theta)\A(\theta)
\end{equation}

With $\alpha^{rs}_{bd}(\theta)$ an invertible matrix. Plugging this in eq.
(\ref{cd1}) and imposing the cancelation of terms of the form
$\B_{g}(\theta^\prime)\A(\theta)$ one obtains:

\begin{eqnarray}
\alpha^{rs}_{bd}(\theta)&=&\alpha(\theta)
\delta^{r}_{b}\delta^{s}_{d}\\
\nonumber
\beta_{bd}(\theta)&=&\beta(\theta)\delta_{bd}\\
\beta(\theta)/\alpha(\theta)&=&-e^{-2\theta}
\s\gamma/\s(2\theta+\gamma)\nonumber
\end{eqnarray}
 
We define the operators:

\begin{eqnarray}
\hat{\D}_{bd}(\theta)&=&\frac{1}{\s 2\theta}[e^{2\theta}\s(2\theta+\gamma)
\D_{bd}(\theta)-\s\gamma\delta_{bd}\A(\theta)]\\
\hat{\B}_{c}(\theta)&=&\frac{\s(2\theta+\gamma)}{\s 2\theta}\;\B_{c}(\theta)
\end{eqnarray}

Now, after some work we  arrive to:

\begin{eqnarray}
\A(\theta)\;\hat{\B}_{c}(\theta^\prime)&=&\frac{\s(\theta+\theta^\prime)
\s(\theta-\theta^\prime-\gamma)}{\s(\theta+\theta^\prime+\gamma)
\s(\theta-\theta^\prime)}\;\hat{\B}_{c}(\theta^\prime)\;\A(\theta)\nonumber\\
&&+\frac{\s\gamma
e^{\theta-\theta^\prime}\s 2\theta}{\s(2\theta+\gamma)
\s(\theta-\theta^\prime)}\;\hat{\B}_{c}(\theta)\;\A(\theta^\prime)\nonumber\\
&&-\frac{\s\gamma
e^{\theta-\theta^\prime}\s 2\theta}{\s(2\theta+\gamma)
\s(\theta+\theta^\prime+\gamma)}\;\hat{\B}_{g}(\theta)\;
\hat{\D}_{gc}(\theta^\prime)
\end{eqnarray}

\begin{eqnarray}
\hat{\D}_{bd}(\theta)\;\hat{\B}_{c}(\theta^\prime)
&=&\frac{\s(\theta+\theta^\prime+2\gamma)\s(\theta-\theta^\prime+\gamma)}
{\s(\theta+\theta^\prime+\gamma)\s(\theta-\theta^\prime)}
\label{comdb}\\
&&R^{(2)}(\theta+\theta^\prime+\gamma)^{eb}_{gh}\;
R^{(2)}(\theta-\theta^\prime)^{ih}_{cd}\;\hat{\B}_{e}(\theta^\prime)\;
\hat{\D}_{gi}(\theta)\nonumber\\
&&-\frac{\s\gamma
e^{\theta-\theta^\prime}\s(2\theta+2\gamma)}{\s(\theta-\theta^\prime)
\s(2\theta+\gamma)}\;R^{(2)}(2\theta+\gamma)^{gb}_{id}
\;\hat{\B}_{g}(\theta)\;\hat{\D}_{ic}(\theta^\prime)\nonumber\\
&&+\frac{\s\gamma e^{\theta-\theta^\prime}\s(2\theta+2\gamma)}
{\s(\theta+\theta^\prime+\gamma)\s(2\theta+\gamma)}\;
R^{(2)}(2\theta+\gamma)^{gb}_{cd}\;\hat{\B}_{g}(\theta)\;\A(\theta^\prime)
\nonumber
\end{eqnarray}

Through the transformation :

\begin{eqnarray}
\theta\rightarrow\theta-\gamma/2\nonumber\\
\theta^\prime\rightarrow\theta^\prime-\gamma/2\nonumber
\end{eqnarray}

Equations (\ref{abba}) and (\ref{dbbd}) follow.\\
It will be also necessary to derive the commutation relations between
the $\hat{\B}$'s. This is obtained using the equality
$M^{11}_{cd}=N^{11}_{cd}$ with $(c,d\geq 2)$. This gives:

\begin{eqnarray}
\hat{\B}_{d}(\theta)\;\hat{\B}_{c}(\theta^{\prime})=
\hat{\B}_{g}(\theta^{\prime})\;
\hat{\B}_{h}(\theta)\;R^{hg}_{cd}(\theta-\theta^{\prime})
\label{bbbb}
\end{eqnarray}
 
\end{section}

\begin{section}{Appendix B : evaluation of $\Delta_{-}(\theta)$ for
SU(n)}

We will work with a chain of length N (remenber
$\theta\rightarrow\theta-\gamma/2$ for the first level). 
One can easily see that:

\begin{eqnarray}
T_{11}(\theta)\1&=&\1\nonumber \\
T_{dd}(\theta)\1&=&\prod^{N}_{i=1}\frac{\s(\theta+\omega_{i}-\gamma/2)}
{\s(\theta+\omega_{i}+\gamma/2)}\1:=\delta_{-}(\theta)\1\nonumber\\
T_{1d}(\theta)\1&\ne&0
\label{tij0}\\ 
T_{ij}(\theta)\1&=&0 \; i\neq j \; i,j\geq 2\nonumber\\
T_{d1}(\theta)\1&=&0\nonumber
\end{eqnarray}

\begin{eqnarray}
\T_{11}(\theta)\1&=&\1\nonumber\\
\T_{dd}(\theta)\1&=&\prod^{N}_{i=1}\frac{\s(\theta-\omega_{i}-\gamma/2)}
{\s(\theta-\omega_{i}+\gamma/2)}\1:=\tilde{\delta}_{-}(\theta)\1\nonumber\\ 
\T_{1d}(\theta)\1&\ne&0
\label{ttij0}\\ 
\T_{ij}(\theta)\1&=&0 \; i\neq j \; i,j\geq 2\nonumber\\
\T_{d1}(\theta)\1&=&0\nonumber
\end{eqnarray}

We will now evaluate the action of $U_{bd}$ on the reference state.

\begin{eqnarray}
\A(\theta)\1&=&T_{1l}(\theta)\T_{l1}(\theta)\1=\nonumber\\
&=&T_{11}(\theta)\T_{11}(\theta)\1\nonumber\\
&=&\1
\end{eqnarray}

\begin{eqnarray}
U_{d1}(\theta)\1&=&T_{dl}(\theta)\T_{l1}(\theta)\1=0\nonumber\\
\end{eqnarray}

Where we have made use of eqs.(\ref{tij0}) and (\ref{ttij0}).
The other elements are not so easy to evaluate. One has to make use of
the following identity which follows by direct calculation:

\begin{equation}
\T_{1d}(\theta)\1=-e^{\gamma}\tilde{\delta}_{-}(\theta)
T_{1d}(-\theta-\gamma)\1
\label{tilde}
\end{equation}

It is also needed to derive from eq.(\ref{yb1}) that:

\begin{eqnarray}
T_{d1}(\theta)T_{1b}(-\theta-\gamma)&=&
T_{1b}(-\theta-\gamma)T_{d1}(\theta)
\label{tdb}\\
&+&\frac{e^{-2\theta}\s\gamma}{\s 2\theta}
[T_{11}(-\theta-\gamma)T_{db}(\theta)-
T_{11}(\theta)T_{db}(-\theta-\gamma)]\nonumber
\end{eqnarray}

We have in addition:

\begin{equation}
U_{db}(\theta)\1=T_{dl}(\theta)\T_{lb}(\theta)\1
\end{equation}

We distinguish now two cases:  $d\neq b$ and $d=b$:\\
(i) $d\neq b$

\begin{eqnarray}
U_{db}(\theta)\1&=&T_{d1}(\theta)\T_{1b}(\theta)\1\\\nonumber
&\propto& T_{d1}(\theta)T_{1b}(-\theta-\gamma)\1=0
\end{eqnarray}

This can be seen  applying both sides of eq.(\ref{tdb}) to the reference
state and using eq.(\ref{tij0}).\\
(ii) $d=b$

\begin{eqnarray}
U_{dd}(\theta)\1&=&T_{d1}(\theta)\T_{1d}(\theta)\1\;+\;\delta_{-}(\theta)
\tilde{\delta}_{-}(\theta)\1\\
&=&\frac{e^{\gamma}}{\s 2\theta}[\,\s(2\theta-\gamma)\tilde{\delta}_{-}
(\theta)\delta_{-}(\theta)+e^{-2\theta}\s\gamma]\1\nonumber
\end{eqnarray}

where we have used eq.(\ref{tdb}) and the fact  that:

\begin{equation}
\tilde{\delta}_{-}(\theta)\delta(-\theta-\gamma)=1
\end{equation}

To conclude, we have:

\begin{eqnarray}
\A(\theta)\1&=&\1\nonumber\\
U_{d1}(\theta)\1&=&0
\label{actr}\\
\hat{\D}_{db}(\theta)\1&=&\Delta_{-}(\theta)\delta_{db}\1\nonumber\\
\hat{\B}_{d}(\theta)\1\;\neq\; 0\nonumber
\end{eqnarray}

Where:

\begin{eqnarray}
\Delta_{-}(\theta)&=&e^{2\theta}\delta_{-}(\theta)\;\tilde{\delta}_{-}(\theta)
\nonumber\\
&=&e^{2\theta}\prod_{i=1}^{N}\frac{\sinh(\theta+\omega_{i}-\gamma/2)
\sinh(\theta-\omega_{i}-\gamma/2)}{\sinh(\theta+\omega_{i}+\gamma/2)
\sinh(\theta-\omega_{i}+\gamma/2)}
\end{eqnarray}

\end{section}

\vspace{3cm}

A.G.R. would like to thank the LPTHE for the kind hospitality, R.
Cuerno for helpful discussions and the Spanish 
M.E.C for financial support under grant AP90 02620085.

\newpage
\section{FIGURE CAPTIONS}
A. Transfer matrix for the open  chain with inhomogeneities
$\tilde{\omega}=(\omega_{N},\ldots,\omega_{1})$.\\
B. Transfer matrix for the periodic chain with inhomogeneities
 $\tilde{\mu}=(\mu_{N},\ldots,\mu_{1})$.\\
C. Identity (\ref{tr}) and its application
to construct the open transfer matrix.\\
D. Representation of the identity (\ref{giro}). Indices are
contracted corresponding with the inhomogeneities.\\
E. Graphical proof of eq.  (\ref{giro}) for  three sites. 
We use the following : 
(a) eq. (\ref{tr}) and $R(0)=1$. (b) eq.
(\ref{yb}), repeat this step for an arbitrary number of sites. (c) eq.
(\ref{norma}). (d)  eq. (\ref{yb}), repeat this step
 for an arbitrary number of sites. (e) eqns. (\ref{norma}) and
(\ref{tr}). (f) eq.  (\ref{yb}). (g) $R(0)=1$.


\begin{thebibliography}{99}
\bibitem{ft} L. D. Faddeev and L. A. Takhtadzhyan,
Russian. Math. Surveys, 34, 11 (1979).
\bibitem{hr}H.J. de Vega Int.J.Mod.Phys. A4, 2371 (1989).
\bibitem{kr} N. Reshetikhin, Theor. Math. Phys. 63, 555 (1986).
\bibitem{mk}H. J. de Vega and M. Karowski, Nucl. Phys. B280, 225 (1987).
\bibitem{krb}N. Reshetikhin and P. B. Wiegman, Phys. Lett. B189, 125 (1987).
\bibitem{sk} E.K.Sklyanin, J.Phys A, 21, 2375 (1988).
\bibitem{fk} A. Foerster and M. Karowski, Berlin preprint, SFB 62/88.
\bibitem{n} N.Reshetikhin, M. Semenov-Tian-Shansky, L.M.P. 19, 133
(1990).
\bibitem{ps} V. Pasquier, H. Saleur, Nucl. Phys. B 330 (1990) 523-556.
\bibitem{qba1}C.Destri and H.J. de Vega, Nucl.Phys B 374(1992) 692-719
\bibitem{ks}H.J. de Vega and A. Gonzalez Ruiz, Boundary K-matrices for
the six vertex and the $n(2n-1)$ $A_{n-1}$ vertex models. LPTHE-PAR
92-45 To appear in J.Phys. A.
\bibitem{mn} L. Mezincescu and R.I. Nepomechie, Int. J. Mod. Phys. A6,
5231(1991).\\
Int. J. Mod. Phys. A7, 5657 (1992).
\bibitem{bq} C.J. Hamer, G.R.W. Quispel and M.T. Batchelor, J. Phys. A
20 (1987) 5677-5693
\bibitem{redu} H. J. de Vega and V. A. Fateev, Int. J. Mod. Phys.
A6, 3221 (1991).
D. Gepner, Caltech preprint.
T. J. Hollowood, Cern/Th. 6888/93, May 1993.
\bibitem{tabrus} Table of Integrals, Series and Products,
I S Gradshteyn and I M Ryzhik, Academic Press, 1980. 

\end{thebibliography}
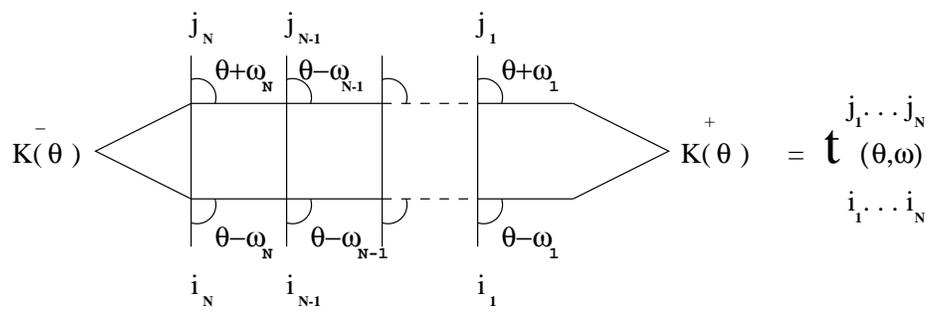
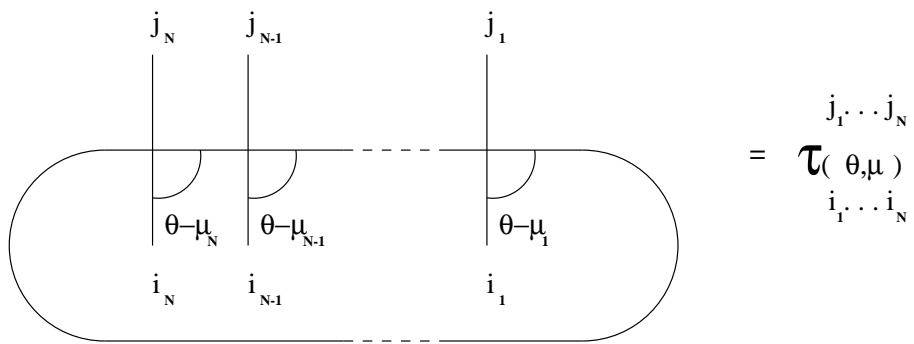
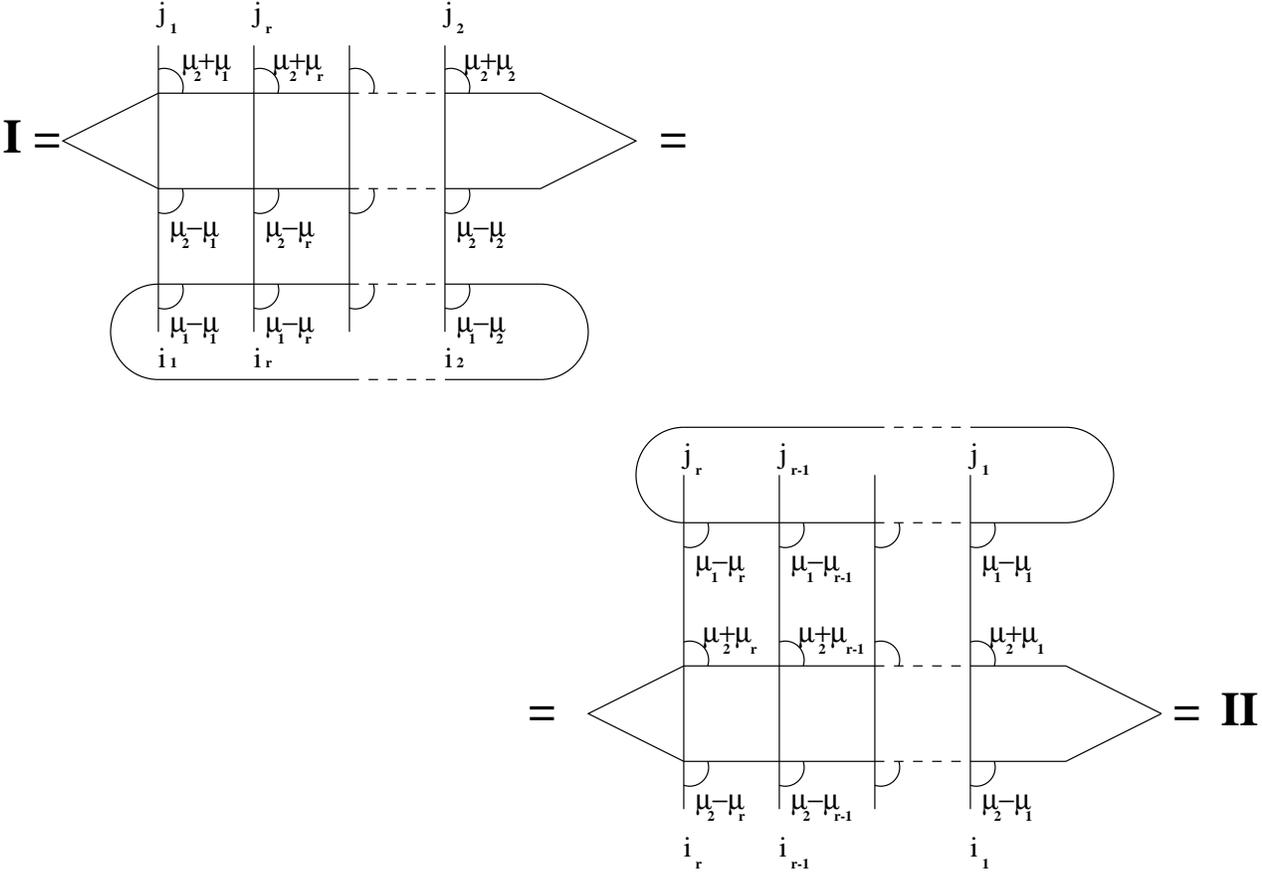
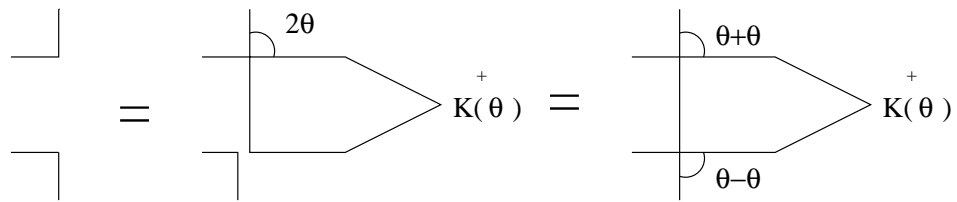
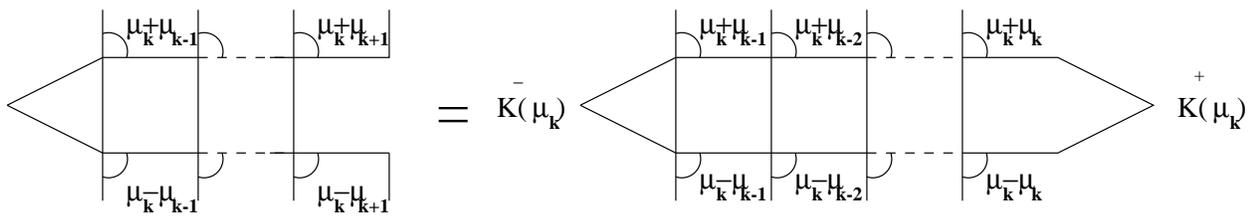
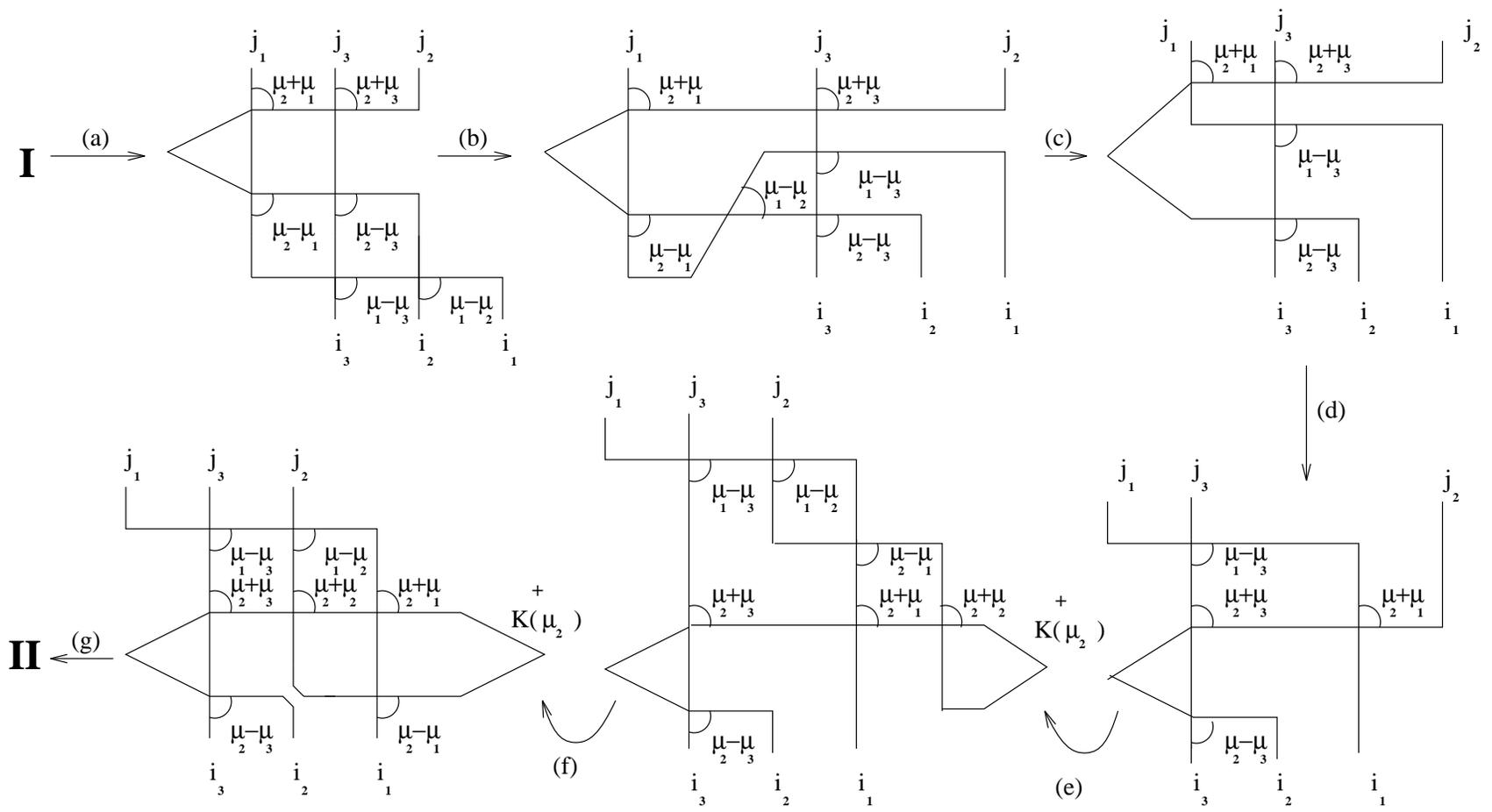
\end{document}